\documentclass[11pt,a4paper]{article}
\usepackage{jheppub}

\usepackage{graphicx}
\usepackage{color}

\usepackage{multirow}

\usepackage{pdflscape}
\usepackage{afterpage}

\usepackage{subcaption}

\usepackage{tikz}
\usetikzlibrary{matrix, positioning}

\def\la{\langle}
\def\ra{\rangle}

\def\spab#1#2{\la #1 \vert \gamma^\mu \vert #2 ]}
\def\spba#1#2{[ #1 \vert \gamma^\mu \vert #2 \ra}

\title{Contribution of third generation quarks to two-loop helicity amplitudes for \boldmath{W} boson pair production in gluon fusion}

\author{Christian Br\o{}nnum-Hansen and}
\author{Chen-Yu Wang}

\affiliation{Institute for Theoretical Particle Physics, KIT, Karlsruhe, Germany}

\emailAdd{christian.broennum-hansen@kit.edu}
\emailAdd{chen-yu.wang@kit.edu}

\abstract{
We compute the contribution of third generation quarks ($t,\ b$) to the two-loop amplitude for on-shell $W$ boson pair production in gluon fusion $gg \to WW$.
We present plots for the amplitude across partonic phase space as well as reference values for two kinematic points.
The master integrals are efficiently evaluated by numerically solving a system of ordinary differential equations.
}

\preprint{\\\\ TTP20-031 \\ P3H-20-045}

\begin{document}

\maketitle
\flushbottom

\section{Introduction}

Production of a $W$ boson pair in gluon fusion, $gg \to WW$, is a loop-induced process.
Although it is expected to be strongly suppressed compared to $q \overline{q} \to WW$, there are two reasons that make it relevant.
First, the large gluon flux at the LHC nearly compensates for the suppression by the strong coupling $\alpha_s$ when compared to quark antiquark annihilation $q \overline{q} \to WW$.
Second, event selection disfavours events with large longitudinal boosts that are due, primarily, to the $q \overline{q} \to WW$ process~\cite{Binoth:2006mf}.
Hence, good understanding of the $gg \to WW$ process is needed for a reliable description of $W$ pair production at the LHC.

The current situation is as follows. One-loop, leading order (LO) cross sections for $W$ pair production in gluon fusion have been computed long ago~\cite{Glover:1988fe,Kao:1990tt,Duhrssen:2005bz,Binoth:2006mf}.
More recently, next-to-leading order (NLO) QCD corrections mediated by massless quark loops were computed in refs.~\cite{Caola:2015psa,Caola:2015rqy} using two-loop amplitudes calculated in refs.~\cite{Caola:2015ila,vonManteuffel:2015msa}.
However, these calculations ignored the contribution of the third quark generation ($t,\ b$) since top quarks cannot be treated as massless.
The goal of this paper is to compute the contribution of third generation quarks to the two-loop amplitude for the $gg \to WW$ process, providing a prerequisite for improved theoretical description of $W$ pair production at the LHC.

Massive fermion propagators that appear when top quark contributions are considered make calculations significantly more demanding compared to the massless case.
Indeed, in the massive case the variety of integrals one has to consider is larger and they are more difficult to compute.

We rely on integration-by-parts (IBP)~\cite{Chetyrkin:1981qh} identities to find linear relations between loop integrals and express the $gg \to WW$ amplitude in terms of a few (master) integrals.
The Laporta algorithm~\cite{Laporta:2001dd} ensures that the system of IBP equations closes.
However, multi-scale integral reductions are computationally expensive and appear to be infeasible for $gg \to WW$ with current publicly available software~\cite{vonManteuffel:2012np,Maierhoefer:2017hyi}.

To overcome this problem, we set the mass of the top quark $m_t$ and the mass of the $W$ boson $m_W$ to integers close to their current experimental values.
Lowering the number of parameters makes the IBP reduction possible.
We have used \texttt{Kira}~\cite{Maierhoefer:2017hyi} for the reduction as well as \texttt{LiteRed}~\cite{Lee:2012cn} and \texttt{Reduze}~\cite{vonManteuffel:2012np} to find symmetry relations between integrals.

We evaluate the master integrals numerically.
A widely used systematic method is that of numerical integration enabled by sector decomposition~\cite{Binoth:2000ps} which, however, is computationally expensive.
Another possibility is to solve numerically a system of differential equations satisfied by the master integrals~\cite{Remiddi:1997ny,Czakon:2020vql}.
Recently, a new method to do this was presented in ref.~\cite{Liu:2017jxz} co-authored by one of the present authors.
This method is particularly suitable for problems with massive particles in the loops.
We use this method to solve a system of differential equations with respect to the $m_{t}^{2}$ variable by moving from $m_{t}^{2} \to -i \infty$ to its physical value.\footnote{To construct the system of differential equations, the integral reduction needs to be parametric in $m_t$. This is however not a bottleneck as the parametric reduction is only required for integrals of low rank.}
The advantage of this method  is that it allows for, essentially, \emph{arbitrary} precision at low computational expense.

The remainder of the paper is organised as follows. Section~\ref{sec:definitions} provides definitions of kinematic variables as well as conventions regarding the $\gamma_5$-matrix and renormalisation of ultraviolet (UV) and infrared (IR) singularities.
In section~\ref{sec:ampcalc} we discuss Feynman diagrams involved in the calculation at one and two loops, the colour structure, and the integral reduction.
A detailed presentation of the numerical approach to the evaluation of the master integrals is given in section~\ref{sec:diffeq}.
In section~\ref{sec:numeval} we evaluate the $gg \to WW$ amplitude at two phase space points and present plots of the amplitude across partonic phase space.
We conclude in section~\ref{sec:conclusions}.

\section{Definitions}\label{sec:definitions}

We study the contribution of third generation quarks ($t$, $b$) to the amplitude of the process
\begin{align}
    \label{equ:process}
    g(p_1) + g(p_2) \to W(p_3) + W(p_4)
    \text{,}
\end{align}
keeping the exact dependence on the top quark mass $m_{t}$ while treating the bottom quark as massless.
We only consider the case where both $W$ bosons couple directly to the quark loop.
Indeed, the single-resonant contribution of an intermediate $Z$ can be ignored as it vanishes for on-shell $W$ pair production~\cite{Binoth:2006mf}.
The process involving an intermediate Higgs boson is known~\cite{Anastasiou:2006hc, Aglietti:2006tp}.

We write the $gg \to WW$ amplitude as
\begin{equation}
    M(\{p_{i}\}, \{\epsilon_{j}\}, m_{t})
    =
    \delta^{a_{1} a_{2}}
    \left( \frac{g_{W}}{\sqrt{2}} \right)^{2}
    A(\{p_{i}\}, \{\epsilon_{j}\}, m_{t}),
\end{equation}
where $a_{1, 2}$ are the colour indices of the external gluons,
$g_{W} = e / \sin \theta_{W}$ is the weak coupling constant,
and $\epsilon_{j}$ are the polarisation vectors of external particles.
We consider the CKM matrix to be an identity matrix.

We set all particles on-shell,
\begin{align}
p_{1}^{2} = p_{2}^{2} = 0,\quad p_{3}^{2} = p_{4}^{2} = m_{W}^{2},
\end{align}
and introduce Mandelstam variables in a standard way
\begin{align}
    s = (p_{1} + p_{2})^{2},\quad
    t  = (p_{1} - p_{3})^{2},\quad
    u = (p_{2} - p_{3})^{2}.
\end{align}
These variables satisfy the relation $s + t + u = 2 m_W^2$.

We decompose the $gg \to WW$ amplitude into 38 tensor structures,
\begin{align}
    A(\{p_{i}\}, \{\epsilon_{j}\}, m_{t})
    &=
    \sum_{I = 1}^{20} A_{I}(s, t, m_{W}, m_{t})\ T_{I}^{\mu \nu} (\{p_{i}\}, \{\epsilon_{j}\}) \epsilon_{3 \mu}^{*}(p_{3}) \epsilon_{4 \nu}^{*}(p_{4}) \nonumber \\
    &\quad + \sum_{I=21}^{38} A_I(s, t, m_W, m_t)\ S_{I}^{\mu \nu} (\{p_{i}\}, \{\epsilon_{j}\}) \epsilon_{3 \mu}^{*}(p_{3}) \epsilon_{4 \nu}^{*}(p_{4}). \label{eq:ampdecomp}
\end{align}
The tensor structures $T_{I}^{\mu \nu}$ are defined in ref.~\cite{vonManteuffel:2015msa} and are parity-even, while $S_{I}^{\mu \nu}$ are parity-odd and are defined in ref.~\cite{Binoth:2006mf}.
Our goal is to calculate the form factors $A_{I}$.

The $Wq\overline{q}$-vertex contains vector and axial parts
\begin{align}
    i \overline{q}_{1} \gamma_{\mu} \frac{1 - \gamma_{5}}{2} q_{2} W^{\mu}.\label{eq:Wqqvertex}
\end{align}
The $gg \to WW$ amplitude can hence be written as a sum of vector-vector, axial-vector, and axial-axial contributions.
The vector-vector and axial-axial terms are parity-even and can be decomposed in terms of tensor structures $T_I^{\mu \nu}$ in eq.~\eqref{eq:ampdecomp}.
On the other hand, the axial-vector term is odd under parity transformations; its decomposition is possible in terms of the tensor structures $S_I^{\mu \nu}$ in eq.~\eqref{eq:ampdecomp}.
If masses of two quarks in a single generation are equal, the parity-odd contribution vanishes~\cite{Glover:1988fe,Kao:1990tt,Binoth:2006mf,Melnikov:2015laa} and it is therefore absent in amplitudes involving only massless quark loops~\cite{Caola:2015ila,vonManteuffel:2015msa}.

To deal with the axial part of the vertex~\eqref{eq:Wqqvertex}, we employ the $\gamma_{5}$-prescription of refs.~\cite{Larin:1993tq, Moch:2015usa} and replace $\gamma_{\mu} \gamma_{5}$ with
\begin{align}
    \gamma_{\mu} \gamma_{5} = - \frac{1}{3 !} \varepsilon_{\mu \nu \rho \sigma} \gamma^{\nu} \gamma^{\rho} \gamma^{\sigma}.
\end{align}
Throughout this paper the Levi-Civita symbol $\varepsilon_{\mu \nu \rho \sigma}$ is defined using the convention of \texttt{FORM} $\varepsilon_{0 1 2 3} = - i$.
Although this $\gamma_{5}$-prescription is much more convenient to work with in dimensional regularisation,
it violates Ward identities of the axial current. To restore the Ward identity, we have to perform a finite renormalisation~\cite{Larin:1991tj},
\begin{equation}
    J^{A}_{\mu}
    = Z_{5} J^{A}_{\mu,\,b} =
    \left[
        1 - \frac{\alpha_{s}}{2 \pi} 2 C_{F} + \mathcal{O}(\alpha_{s}^{2})
    \right]
    J^{A}_{\mu,\,b}
    \text{,}
\end{equation}
where $J^{A}$ and $J^{A}_{b}$ stand for renormalised and unrenormalised axial currents respectively.

\subsection{Pole structure}
\label{sec:polestructure}

Throughout the calculation we employ dimensional regularisation and set the space-time dimensionality to $d = 4 - 2 \epsilon$.
The singularities ubiquitous in loop calculations appear as poles in $\epsilon$ of either ultraviolet (UV) or infrared (IR) origin.
UV poles are absorbed through the introduction of renormalisation factors leading to a renormalised amplitude.

Expanding the unrenormalised amplitude $A_{b}$ in the bare strong coupling $\alpha_{s}^{0}$ we have
\begin{align}
A_{b} = A_{b}^{(0)} + \frac{\alpha_{s}^{0}}{2 \pi} A_{b}^{(1)} + \left( \frac{\alpha_{s}^{0}}{2 \pi} \right)^{2} A_{b}^{(2)} + \mathcal{O}\left( (\alpha_{s}^{0})^{3} \right).
\end{align}
Following~\cite{Chen:2017jvi, Behring:2019oci} we employ a hybrid renormalisation scheme where the gluon field $G_{\mu}$ and the top quark mass $m_{t}$ are in the on-shell scheme while the strong coupling constant $\alpha_{s}$ is renormalised in the $\overline{\text{MS}}$ scheme.
We have
\begin{align}
\label{equ:renormalisationconstants}
\alpha_{s}^{0} = \mu^{2 \epsilon} S_{\epsilon} Z_{\alpha_{s}} \alpha_{s},
\qquad
G^{0}_{\mu} = \sqrt{Z_{g}} G_{\mu},
\qquad
m_{t}^{0} = Z_{m_{t}} m_{t},
\end{align}
where $\mu$ is the renormalisation scale and $S_{\epsilon} = (4 \pi)^{- \epsilon} e^{\epsilon \gamma_{E}}$.
The renormalisation constants are expanded in the coupling
\begin{align}
Z = \sum_{n = 0} \left( \frac{\alpha_{s}}{2 \pi} \right)^{n} Z^{(n)},
\qquad
Z^{(0)} = 1.
\end{align}
The renormalised amplitude is related to the unrenormalised one by
\begin{align}
\label{equ:renormalisedamplitude}
A (\epsilon, \mu, \alpha_{s}, m_{t})
& =
Z_{g} A_{b} (\epsilon, \alpha_{s}^{0}, m_{t}^{0})
=
\frac{\alpha_{s}}{2 \pi}
A^{(1)}(\epsilon, m_{t})
+
\left( \frac{\alpha_{s}}{2 \pi} \right)^{2}
A^{(2)}(\epsilon, m_{t})
+
\mathcal{O}(\alpha_{s}^{3}),
\\
A^{(1)}(\epsilon, m_{t})
& =
\mu^{2 \epsilon} S_{\epsilon}
A_{b}^{(1)} (\epsilon, m_{t}),
\\
A^{(2)}(\epsilon, m_{t})
& =
\mu^{2 \epsilon} S_{\epsilon}
\left[
(Z_{g}^{(1)} + Z_{\alpha_{s}}^{(1)})
A_{b}^{(1)} (\epsilon, m_{t})
+
m_{t} Z^{(1)}_{m_{t}}
C_{b}^{(1)} (\epsilon, m_{t})
\right]
\nonumber
\\
& \phantom{= {}}
+
\left( \mu^{2 \epsilon} S_{\epsilon} \right)^{2}
A_{b}^{(2)} (\epsilon, m_{t}),\label{eq:A2ren}
\end{align}
where we used the fact that the tree level amplitude $A_b^{(0)}$ vanishes.
Note that while the renormalisation factors for the strong coupling and the gluon wave function are multiplicative at the level of the amplitude,
the factor for the mass renormalisation is not.
For this reason, the mass counterterm $C^{(1)}$ is calculated separately.

The relevant renormalisation factors are \cite{Gross:1973id, Politzer:1973fx, Melnikov:2000zc, Beenakker:2002nc, Czakon:2007wk}
\begin{align}
Z_{\alpha_s}^{(1)}
& =
- \frac{\gamma_{g}(n_{l} + 1)}{\epsilon},
\\
Z_{g}^{(1)}
& =
S_{\epsilon}
\left( \frac{4 \pi \mu^{2}}{m_{t}^{2}} \right)^{\epsilon}
\Gamma(1 + \epsilon)\,
T_{F}
\left[ - \frac{2}{3 \epsilon} \right]
,
\\
Z_{m_{t}}^{(1)}
& =
S_{\epsilon}
\left( \frac{4 \pi \mu^{2}}{m_{t}^{2}} \right)^{\epsilon}
\Gamma(1 + \epsilon)\,
C_{F}
\left[ - \frac{3}{2 \epsilon} - \frac{2}{1 - 2 \epsilon} \right]
,
\end{align}
where $\gamma_{g}(n_{l} + 1) = \frac{11}{6} C_{A} - \frac{2}{3} T_{F} (n_{l} + 1)$ and $n_{l}$ is the number of massless fermions.
The divergent contribution of the massive quark flavour in $\gamma_{g}(n_{l} + 1)$ cancels between the gluon field and the coupling renormalisation,
while the light quark contributions are absorbed into Catani's operator, as explained below.

In general, IR poles of loop amplitudes cancel against contributions of real emission processes.
In the virtual amplitudes the IR singularities factorise in a universal manner~\cite{Catani:1998bh}, this allows us to write the amplitude as a sum of IR poles and a finite remainder.

Since $gg \to WW$ vanishes at tree level, the pole structure of the two-loop amplitude is particularly simple.
Furthermore, the amplitude is a singlet in colour space.
Hence, we can write the renormalised amplitude as
\begin{align}
    \label{eq:irfactorisation}
    A^{(2)}(\epsilon, \mu)
    =
    \boldsymbol{I}^{(1)} (\epsilon, \mu) A^{(1)}(\epsilon, \mu) + F^{(2)}(\epsilon, \mu),
\end{align}
where $F^{(2)}$ is the finite remainder. The Catani operator reads
\begin{align}
    \boldsymbol{I}^{(1)} (\epsilon, \mu)
    & =
    -N(\epsilon)
    \left( \frac{C_{A}}{\epsilon^{2}} + \frac{\gamma_{g}(n_{l})}{\epsilon} \right)
    \left( \frac{\mu^{2} e^{-i \pi}}{s} \right)^{\epsilon},
\end{align}
where $N(\epsilon) = e^{\epsilon \gamma_{E}} / \Gamma (1 - \epsilon)$ and
$\gamma_{g}(n_{l}) = \frac{11}{6} C_{A} - \frac{2}{3} T_{F} n_{l}$.
In the following we will set $T_{F} = \frac{1}{2}$.
In order to obtain the finite remainder of the two-loop amplitude, we require the one-loop amplitude, $A^{(1)}$, expanded through $\mathcal{O}(\epsilon^{2})$.

\section{Amplitude calculation}\label{sec:ampcalc}

In this section we discuss the calculation of the amplitudes $A^{(1)}$, $C^{(1)}$, and $A^{(2)}$.

\subsection{One loop}

We generate 8 one-loop diagrams using \texttt{QGRAF}~\cite{Nogueira:1991ex} and perform colour and Dirac algebra as well as projection of the form factors shown in eq.~\eqref{eq:ampdecomp} using \texttt{FORM}~\cite{Vermaseren:2000nd,Kuipers:2013pba,Ruijl:2017dtg}.
Two triangle diagrams with $g^\star \to WW$ transition vanish due to colour conservation.
The remaining six box diagrams can be mapped to 5 independent topologies.
We perform the integral reduction step using IBP identities,
and express the form factors as linear combinations of 16 master integrals.
The form factors for the mass counterterm amplitude $C^{(1)}$ are computed in a similar fashion.
The one-loop amplitudes were checked against \texttt{FeynArts} and \texttt{FeynCalc}~\cite{Hahn:2000kx, Mertig:1990an, Shtabovenko:2016sxi, Shtabovenko:2020gxv}.

\subsection{Two loops}

\begin{figure}[ht]
    \centering
    \begin{subfigure}[ht]{0.23\linewidth}
        \centering
        \includegraphics[width=0.9\linewidth]{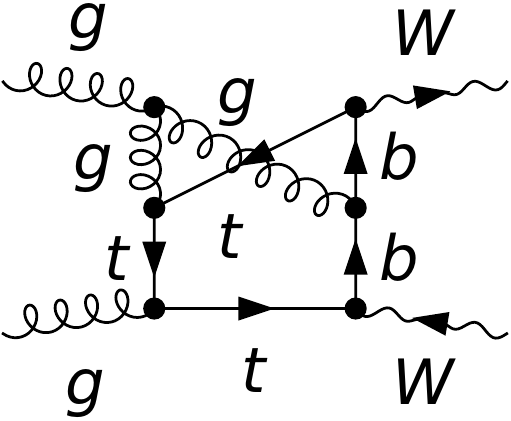}
    \end{subfigure}
    ~
    \begin{subfigure}[ht]{0.23\linewidth}
        \centering
        \includegraphics[width=0.9\linewidth]{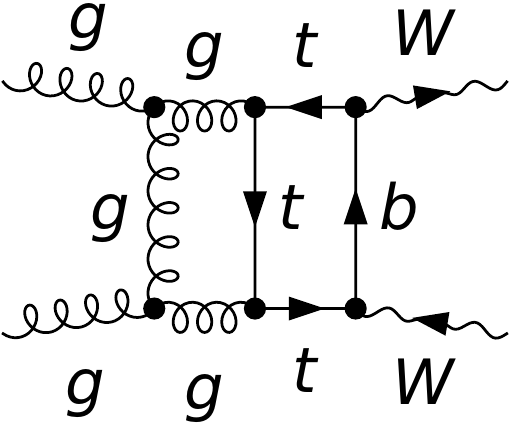}
    \end{subfigure}
    ~
    \begin{subfigure}[ht]{0.23\linewidth}
        \centering
        \includegraphics[width=0.9\linewidth]{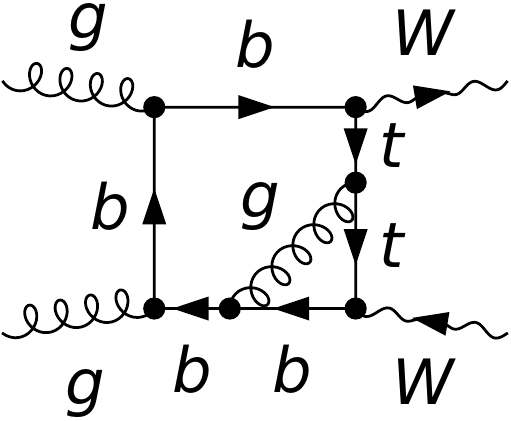}
    \end{subfigure}
    ~
    \begin{subfigure}[ht]{0.23\linewidth}
        \centering
        \includegraphics[width=0.9\linewidth]{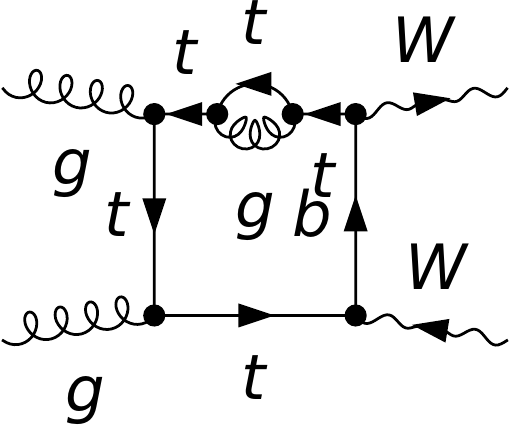}
    \end{subfigure}
    \caption{Representative two-loop Feynman diagrams for $gg \to WW$.}
    \label{fig:diagrams}
\end{figure}

At two loops we generate 136 diagrams using the same steps as above.
To test these steps of our implementation, a subset of unreduced diagrams were crosschecked numerically against \texttt{FeynArts} and \texttt{FeynCalc}.
In figure~\ref{fig:diagrams} we show a few representative diagrams that contribute to the two-loop $gg \to WW$ amplitude.

\begin{figure}
    \centering
    \includegraphics[width=0.8\linewidth]{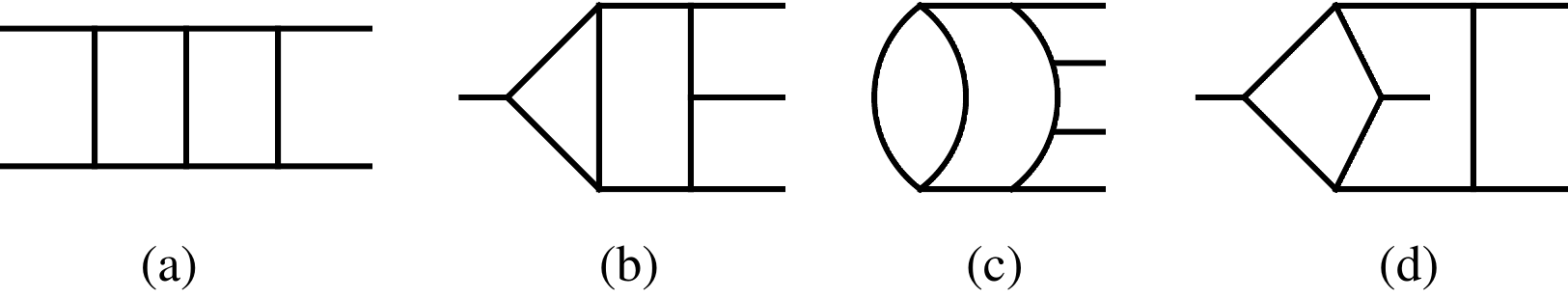}
    \caption{The four basic topologies. The internal lines can be both massive and massless.
        The first three (a)--(c) are planar and can at most have 5 massive internal lines,
        while the nonplanar topology (d) can at most have 4 massive internal lines.}
    \label{fig:basictopologies}
\end{figure}

Since the weak bosons are not colour charged, the colour structure follows that of the gluon self-energy two-loop diagrams with a closed quark loop.
These diagrams are given in table~\ref{tab:colour}. All 136 diagrams contributing to $gg \to WW$ can be formed by attaching two $W$ bosons to the quark loop and by pinching propagators.
This results in four basic topologies shown in figure~\ref{fig:basictopologies}. It is clear from this classification that only class L has nonplanar contributions.
We find 33 non-vanishing diagrams in class L (of which 17 are nonplanar), 20 in class S and 40 in LS.

This classification motivates splitting the amplitude into leading ($N_{C}$) and sub-leading ($1 / N_{C}$) colour contributions,
\begin{equation}
    A^{(2)}
    =
    N_{c} A^{(2), [1]}
    +
    \frac{1}{N_{c}} A^{(2), [-1]}.
\end{equation}
Note that both $A^{(2), [1]}$ and $A^{(2), [-1]}$ are gauge invariant.
We also observe that $A^{(2), [-1]}$ is finite after mass renormalisation and has no infrared poles.

\begin{table}
    \centering
    \begin{tabular}{|c|c|c|}
        \hline \hline
        & \textbf{Class} & \textbf{Colour factor} \\
        \hline \hline
        \multirow{6}{*}{\includegraphics[width=0.15\linewidth]{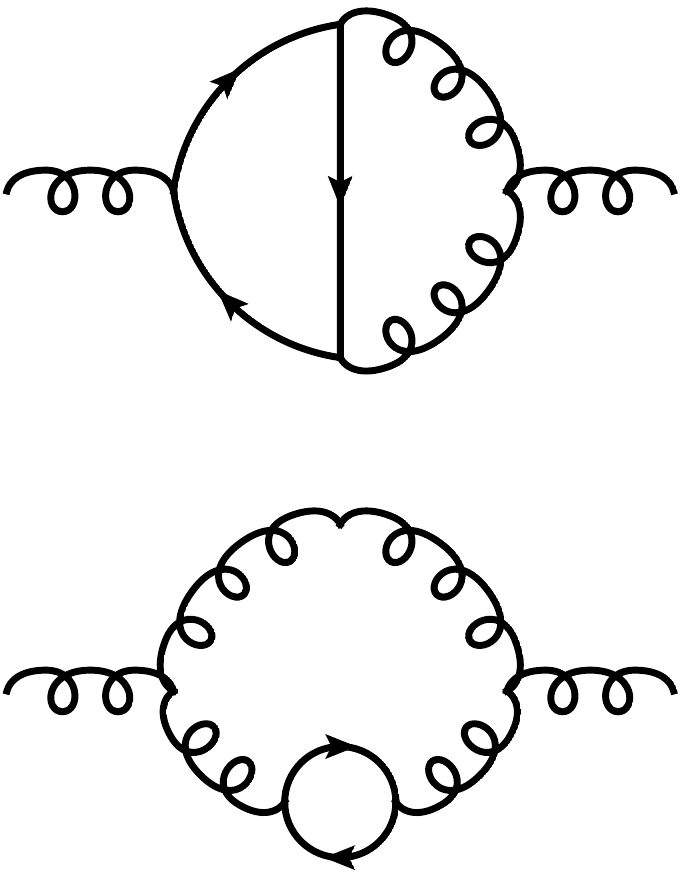}} & & \\
        & & \\
        & L & $N_C$ \\
        & & \\
        & & \\
        & & \\
        \hline
        \multirow{3}{*}{\includegraphics[width=0.15\linewidth]{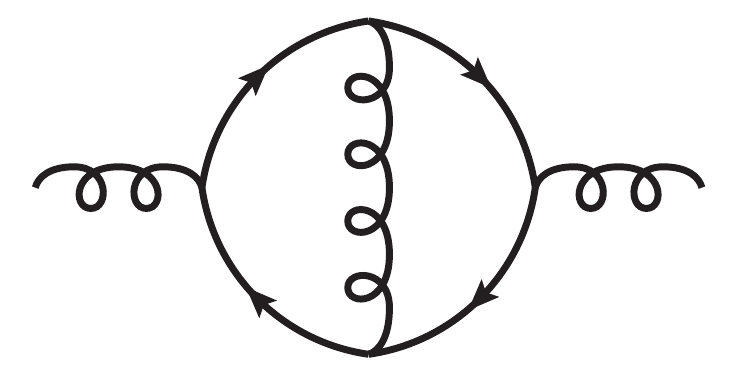}} & & \\
        & S & $\frac{1}{N_C}$ \\
        & & \\
        \hline
        \multirow{3}{*}{\includegraphics[width=0.15\linewidth]{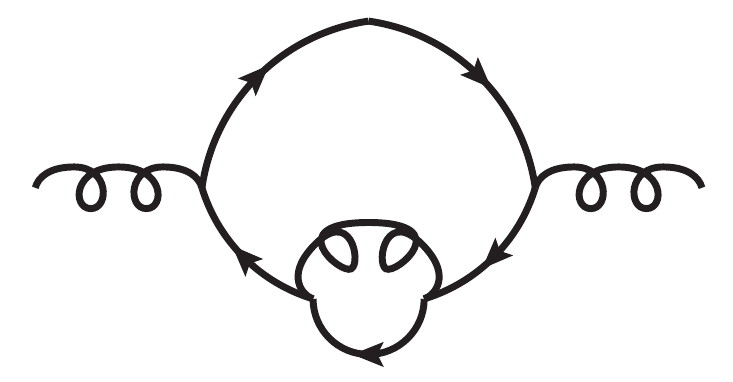}} & & \\
        & LS & $\left( N_C - \frac{1}{N_C} \right)$ \\
        & & \\
        \hline \hline
    \end{tabular}
    \caption{The classification of diagrams in colour structures follows that of the two-loop gluon self-energy diagrams involving a closed fermion loop. This motivates splitting the amplitude into leading and sub-leading colour.}
    \label{tab:colour}
\end{table}

In order to perform an IBP reduction we express the amplitude in terms of integral families.
An integral family is a set of propagators and irreducible scalar products (ISPs) that forms a basis of the linear space spanned by all scalar products containing loop momenta.
For four-point kinematics in four dimensions there are 9 independent scalar products at two loops.
Before the integral reduction the form factors can be written as
\begin{equation}
    A_{I}^{(2)}
    =
    \sum_{T = 1}^{N_{T}} \sum_{\vec{a}_T}
    c_{I T \vec{a}_T}^{(2)} I_{T} (\vec{a}_T).\label{eq:unreducedamp}
\end{equation}
In eq.~\eqref{eq:unreducedamp} $N_{T}$ is the number of independent integral families and $\vec{a}_{T} = (a_{1}, \ldots, a_{9})$.
Each $a_{i}$ is an integer power of the 9 independent scalar products in the integral family $T$.
Using symmetries between diagrams, including crossing symmetry,
we find a total of $N_{T} = 35$ families of type (a), (b), and (d). The coefficients of integrals, $c_{I T \vec{a}_T}^{(2)}$, are rational functions of $s$, $t$, $m_{t}$, $m_{W}$ and space-time dimensionality $d$.

\begin{figure}[ht]
    \centering
    \begin{subfigure}[ht]{0.3\linewidth}
        \centering
        \includegraphics[width=0.8\linewidth]{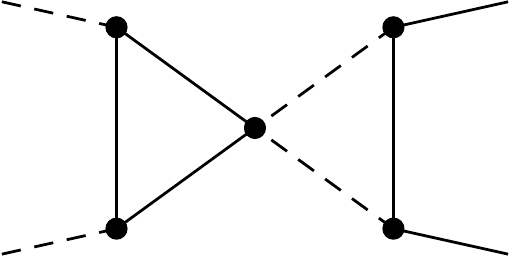}
        \caption{one-loop squared}
    \end{subfigure}
    ~
    \begin{subfigure}[ht]{0.3\linewidth}
        \centering
        \includegraphics[width=0.8\linewidth]{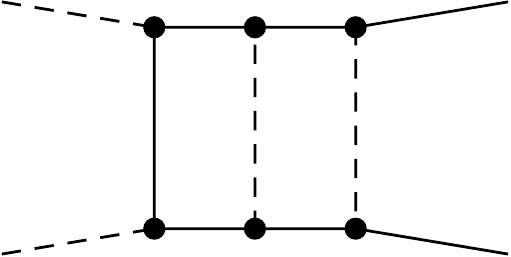}
        \caption{planar no.\ 1}
    \end{subfigure}
    ~
    \begin{subfigure}[ht]{0.3\linewidth}
        \centering
        \includegraphics[width=0.8\linewidth]{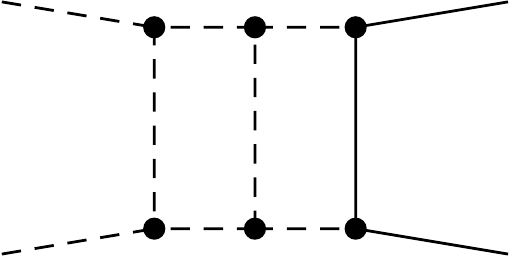}
        \caption{planar no.\ 2}
    \end{subfigure}
    \\
    \begin{subfigure}[ht]{0.3\linewidth}
        \centering
        \includegraphics[width=0.8\linewidth]{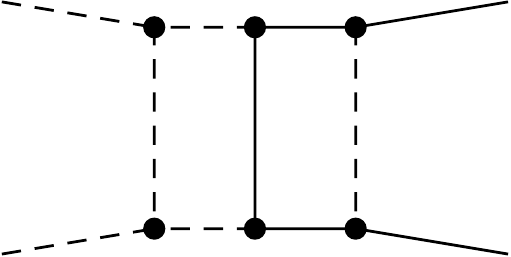}
        \caption{planar no.\ 3}
    \end{subfigure}
    ~
    \begin{subfigure}[ht]{0.3\linewidth}
        \centering
        \includegraphics[width=0.8\linewidth]{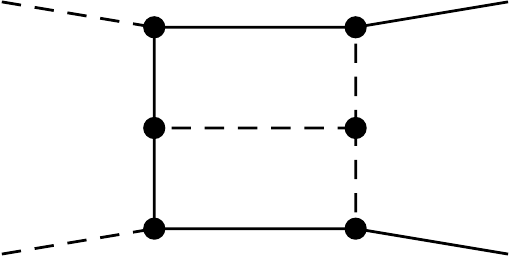}
        \caption{planar no.\ 4}
    \end{subfigure}
    ~
    \begin{subfigure}[ht]{0.3\linewidth}
        \centering
        \includegraphics[width=0.8\linewidth]{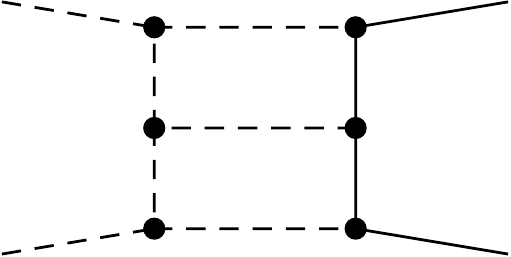}
        \caption{planar no.\ 5}
    \end{subfigure}
    \\
    \begin{subfigure}[ht]{0.3\linewidth}
        \centering
        \includegraphics[width=0.8\linewidth]{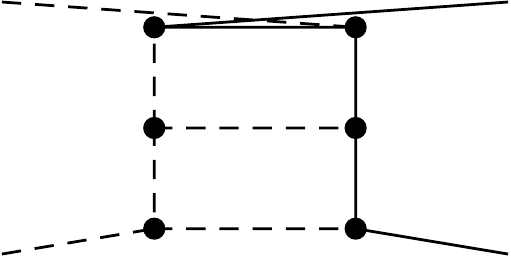}
        \caption{planar no.\ 6}
    \end{subfigure}
    ~
    \begin{subfigure}[ht]{0.3\linewidth}
        \centering
        \includegraphics[width=0.8\linewidth]{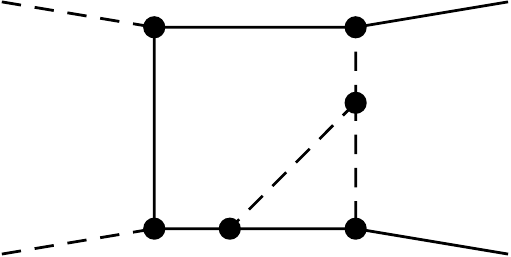}
        \caption{planar no.\ 7}
    \end{subfigure}
    ~
    \begin{subfigure}[ht]{0.3\linewidth}
        \centering
        \includegraphics[width=0.8\linewidth]{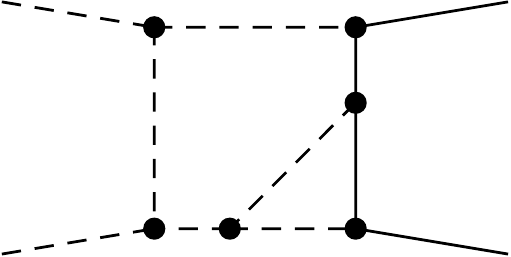}
        \caption{planar no.\ 8}
    \end{subfigure}
    \\
    \begin{subfigure}[ht]{0.3\linewidth}
        \centering
        \includegraphics[width=0.8\linewidth]{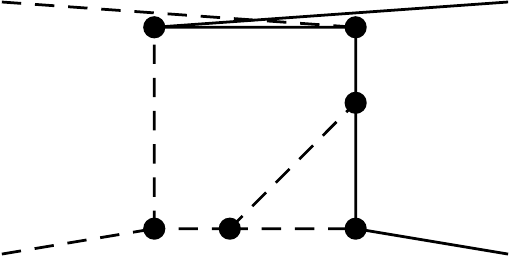}
        \caption{planar no.\ 9}
    \end{subfigure}
    ~
    \begin{subfigure}[ht]{0.3\linewidth}
        \centering
        \includegraphics[width=0.8\linewidth]{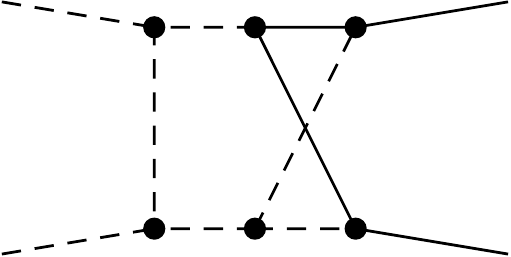}
        \caption{nonplanar no.\ 1}
    \end{subfigure}
    ~
    \begin{subfigure}[ht]{0.3\linewidth}
        \centering
        \includegraphics[width=0.8\linewidth]{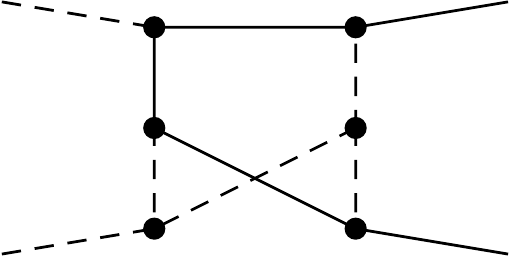}
        \caption{nonplanar no.\ 2}
    \end{subfigure}
    \\
    \begin{subfigure}[ht]{0.3\linewidth}
        \centering
        \includegraphics[width=0.8\linewidth]{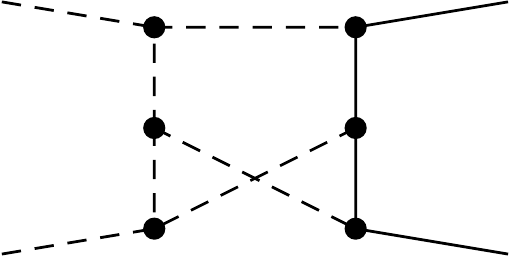}
        \caption{nonplanar no.\ 3}
    \end{subfigure}
    ~
    \begin{subfigure}[ht]{0.3\linewidth}
        \centering
        \includegraphics[width=0.8\linewidth]{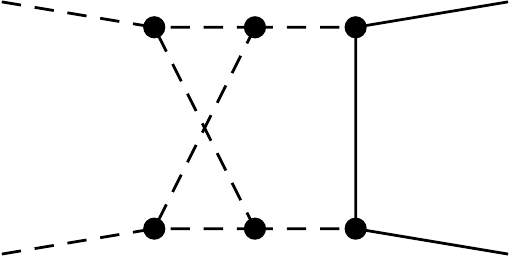}
        \caption{nonplanar no.\ 4}
    \end{subfigure}
    ~
    \begin{subfigure}[ht]{0.3\linewidth}
        \centering
        \includegraphics[width=0.8\linewidth]{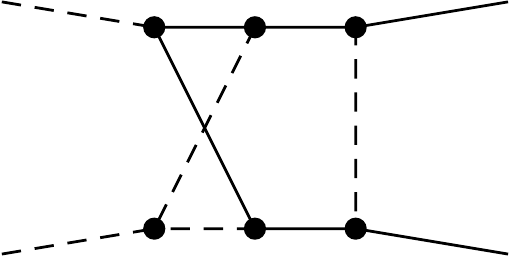}
        \caption{nonplanar no.\ 5}
    \end{subfigure}
    \caption{Topologies of integral families.
        Solid and dashed lines correspond to massive and massless particles respectively.
        Internal massive particles have mass $m_{t}$ while external massive particles have mass $m_{W}$.
    	All nine planar topologies and nonplanar no.\ 2 and 3 can be crossed ($p_1 \leftrightarrow p_2$) giving a total of 26 topologies.}
    \label{fig:family}
\end{figure}

The 35 families can be further reduced to 25 two-loop irreducible families and a single one-loop squared family using IBP identities.
Family definitions are given in appendix~\ref{sec:families} and the corresponding topologies are shown in figure~\ref{fig:family}.
We use \texttt{Kira} \cite{Maierhoefer:2017hyi} to reduce all integrals $I_{T}(\vec{a}_T)$ appearing in the scattering amplitude
to a set of 334 master integrals.

To complete the integral reduction using reasonable time and resources,
several measures have been taken.
First, the integrals depend on the kinematic variables $s$ and $t$,
the masses $m_{t}$ and $m_{W}$,
and space-time dimensionality $d$.
Keeping all of them parametric makes the reduction formidably complicated.
To overcome this problem,
we set $m_{t} = 173 \text{~GeV}$ and $m_{W} = 80 \text{~GeV}$ in the IBP reduction,
keeping only $s$, $t$, and $d$ as parameters.\footnote{Although the numerical values of $m_{t}$ and $m_{W}$ that we use are different from current experimental values by about $0.1\%$ and $0.5\%$ respectively \cite{Zyla:2020zbs},
the impact of these differences on the two-loop amplitude is negligible.
If needed, these differences can be taken into account by Taylor expanding the (unreduced) amplitude in terms of the mass differences and using the same set of reduction tables and master integrals to compute the correction.}
This simplifies the reduction tables and cuts down the run times considerably.

Second, the size of reduction tables can be reduced further by a careful choice of master integrals.
Our guiding principle in choosing master integrals is absence of denominators with non-factorisable dependence on kinematic variables and space-time dimensionality $d$ as well as avoiding denominators that lead to poles at non-integer values of $d$ (e.g.\ $3 d - 10$)~\cite{Smirnov:2020quc, Usovitsch:2020jrk}.
Furthermore, we aim at having simple differential equations for fast numerical evaluation.
By trial and error, we find that master integrals with at most one squared propagator or a single irreducible scalar product are sufficient to satisfy the above requirements.
In a few sectors integrals with two squared propagators are needed, but this appears to be an exception rather than the rule.

Third,
we use the \texttt{select\_masters\_reduction} feature implemented in \texttt{Kira}
to project integrals onto one master integral at a time.
Our final reduction tables are of the order of hundred MB each,
which makes applying reduction rules to the amplitude a manageable job.

\section{Differential equation}\label{sec:diffeq}

Having expressed the full amplitude through master integrals,
we need to evaluate them.
The master integrals are defined as follows
\begin{equation}
    I(a_{1}, \ldots, a_{9})
    =
    \int \left( \prod_{n = 1}^{2} e^{\epsilon \gamma_{E}} \frac{\mathrm{d}^{d} l_{n}}{i \pi^{d / 2}} \right)
    \frac{1}{D_{1}^{a_{1}} D_{2}^{a_{2}} \cdots D_{9}^{a_{9}}}
    \text{,}
\end{equation}
where $D_{i}$ are denominators in one of the 26 families given in appendix~\ref{sec:families}.
Their topologies are shown in figure~\ref{fig:family}.
Note that we absorb a factor of $- i (4 \pi)^{2 - \epsilon} e^{\epsilon \gamma_{E}}$ per loop into the definition of master integrals.

To evaluate all 334 two-loop master integrals with massive propagators,
we employ the imaginary mass method proposed in ref.~\cite{Liu:2017jxz}.
In the original formulation of this method,
an imaginary mass term $- i \eta$ is added to all propagators. The differential equation with respect to $\eta$ is solved numerically
starting at $\eta \to \infty$ and progressing towards a physical point $\eta = 0^{+}$.
The boundary conditions involve vacuum integrals with mass $m^{2} = - i \eta$,
as no physical masses or kinematic variables survive in this limit.
Adding $- i \eta$ to massless propagators alters the behaviour of integrals near the physical point $\eta = 0^{+}$ and generates a singularity in the differential equation.
In order to compute the result at the physical point it is then necessary to fix constants in a formal solution of the differential equation, by matching against another point within its radius of convergence.

We employ a variant of the original method, where the imaginary mass parameter $- i \eta$ is introduced only to the massive propagators.
Setting the mass $m_{t}^{2}$ of massive propagators to $m_{t}^{2} - i \eta$ requires no extra work as far as IBP reductions are concerned.\footnote{We need to keep $m_{t}$ as a parameter when constructing differential equations using IBP identities, but this is unproblematic since the integrals that appear in the differential equations are much simpler than those in the amplitude.}
Since our diagrams already have many massive propagators,
the boundary conditions at $\eta \to \infty$ are remarkably simple,
with only 5 planar integrals and 1 nonplanar, massless 3-point integral needed to fix all master integrals at the boundary.
The topologies of the boundary integrals are shown in figure~\ref{fig:boundary},
see appendix~\ref{sec:boundary} for their definitions~\cite{tHooft:1978jhc, Chetyrkin:1980pr, Scharf:1993ds, Gehrmann:1999as, Gehrmann:2005pd}.

\begin{figure}[ht]
    \centering
    \begin{tikzpicture}
        \matrix (mat) [left delimiter=\lbrace]
        {
            \node {region 1:}; &
            \node (r1a) {$m_{t}^{-6 - 4 \epsilon} \times$};
            \node (r1b) [right=0 of r1a] {\includegraphics[width=0.2\linewidth]{figures/boundary_i_2.pdf}};
            \\
            \node {region 2:}; &
            \node (r2a) {$m_{t}^{-4 - 2 \epsilon} \times$};
            \node (r2b) [right=0 of r2a] {\includegraphics[width=0.2\linewidth]{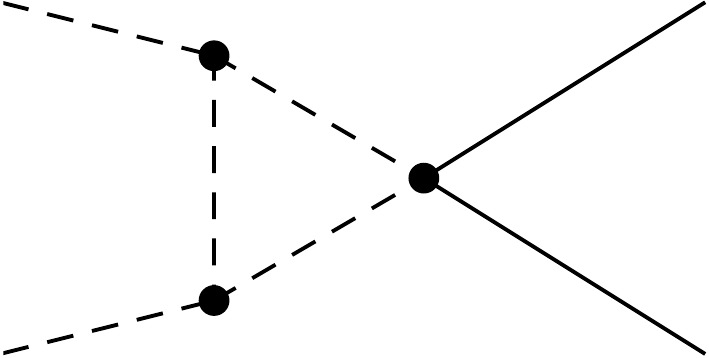}};
            \node (r2c) [right=0 of r2b] {$\times$};
            \node (r2d) [right=0 of r2c] {\includegraphics[width=0.2\linewidth]{figures/boundary_i_1.pdf}};
            \\
            \node {region 3:}; &
            \node (r3a) {$m_{t}^{-2} \times$};
            \node (r3b) [right=0 of r3a] {\includegraphics[width=0.2\linewidth]{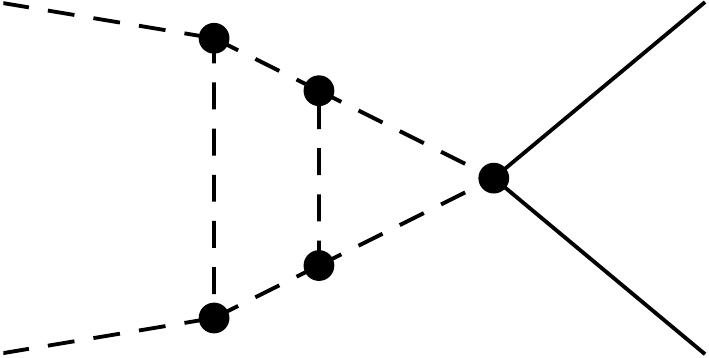}};
            \\
        };
        \node (propto) [left=15pt of mat] {$\propto$};
        \node (diagram) [left=5pt of propto] {\includegraphics[width=0.2\linewidth]{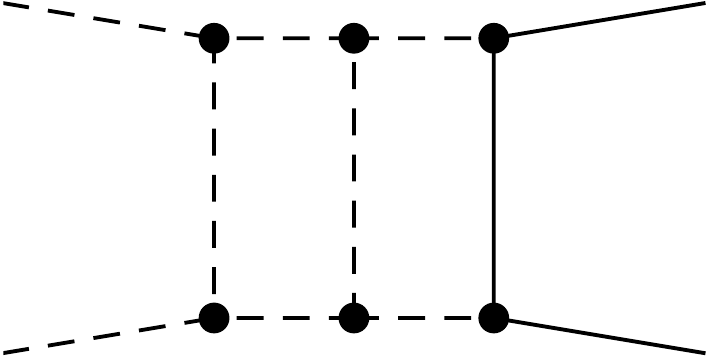}};
    \end{tikzpicture}
    \caption{A typical master integral and its leading regions.
        Solid and dashed lines correspond to massive and massless particles respectively.
        Internal massive particles have mass $m_{t}$, while external massive particles have mass $m_{W}$.}
    \label{fig:boundary_example}
\end{figure}

By constructing a differential equation with respect to $m_{t}^{2}$ and
choosing the boundary condition at $m_{t}^{2} \to - i \infty$,
the differential equation can be used to evaluate master integrals at the physical mass, $m_{t}^{2} = (173 \text{~GeV})^{2}$.
At $\eta \to \infty$, the master integrals receive contributions from 3 regions:
\begin{enumerate}
    \item All internal momenta are comparable to $m_{t}^{2} \to -i \infty$.
    \item Some internal momenta that form a closed loop are comparable to $m_{t}^{2} \to -i \infty$,
        while the remaining momenta are much smaller than $m_{t}^{2}$
        and are comparable to other kinematic parameters (i.e.\ $s$, $t$, $m_{W}^{2}$).
    \item All internal momenta are much smaller than $m_{t}^{2} \to -i \infty$.
\end{enumerate}
In figure~\ref{fig:boundary_example} we show a typical master integral and its regions.
The boundary condition in each region can be expressed in terms of the boundary integrals given in figure~\ref{fig:boundary} through IBP reduction,
together with an overall scaling factor of $m_{t}$.

Note that if we add $- i \eta$ to all propagators, only the first region contributes to the integrals.
In general, the fewer propagators one takes to be infinitely massive,
the more complicated boundary conditions one needs to consider.
However, since we change the original integrals less,
the singularity at the physical point, $\eta = 0^{+}$, is simpler.
In fact, for the problem at hand, the differential equation is \emph{regular} at the physical point and no additional complications arise.

\begin{figure}[ht]
    \centering
    \begin{subfigure}[ht]{0.3\linewidth}
        \centering
        \includegraphics[width=0.8\linewidth]{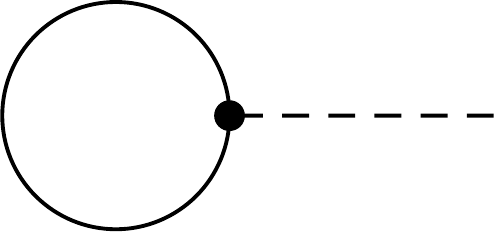}
        \caption{$I_{1}$}
    \end{subfigure}
    ~
    \begin{subfigure}[ht]{0.3\linewidth}
        \centering
        \includegraphics[width=0.8\linewidth]{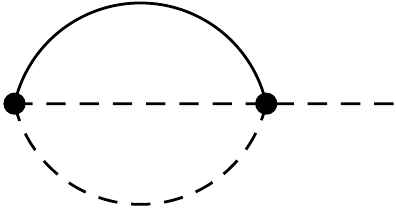}
        \caption{$I_{2}$}
    \end{subfigure}
    ~
    \begin{subfigure}[ht]{0.3\linewidth}
        \centering
        \includegraphics[width=0.8\linewidth]{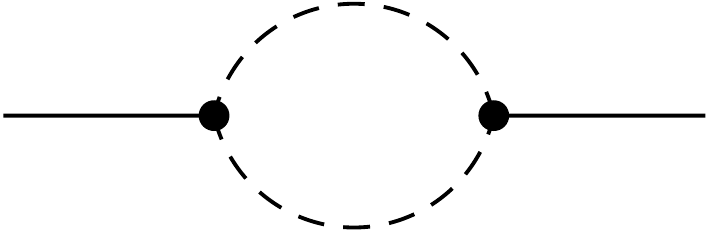}
        \caption{$I_{3}$}
    \end{subfigure}
    \\
    \begin{subfigure}[ht]{0.3\linewidth}
        \centering
        \includegraphics[width=0.8\linewidth]{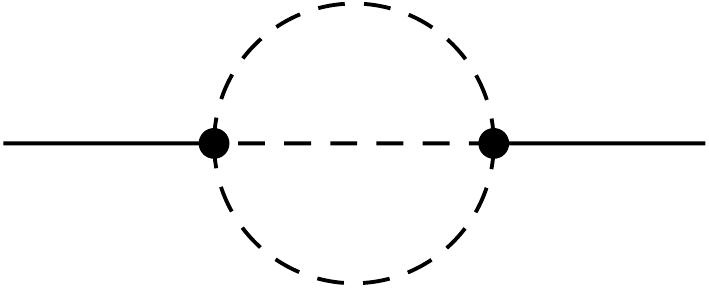}
        \caption{$I_{4}$}
    \end{subfigure}
    ~
    \begin{subfigure}[ht]{0.3\linewidth}
        \centering
        \includegraphics[width=0.8\linewidth]{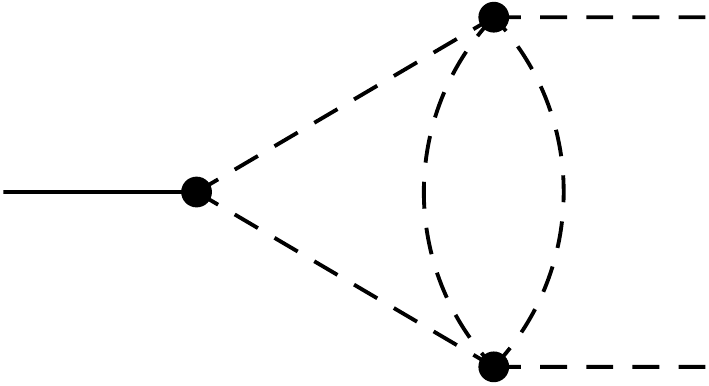}
        \caption{$I_{5}$}
    \end{subfigure}
    ~
    \begin{subfigure}[ht]{0.3\linewidth}
        \centering
        \includegraphics[width=0.8\linewidth]{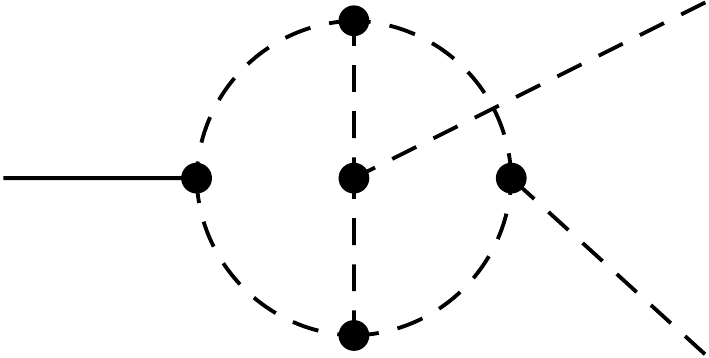}
        \caption{$I_{6}$}
    \end{subfigure}
    \caption{Topologies of boundary integrals.
        Solid and dashed lines correspond to massive and massless particles respectively.
        See appendix~\ref{sec:boundary} for their explicit expressions.
        }
    \label{fig:boundary}
\end{figure}

\begin{figure}[ht]
    \centering
    \includegraphics[width=\linewidth]{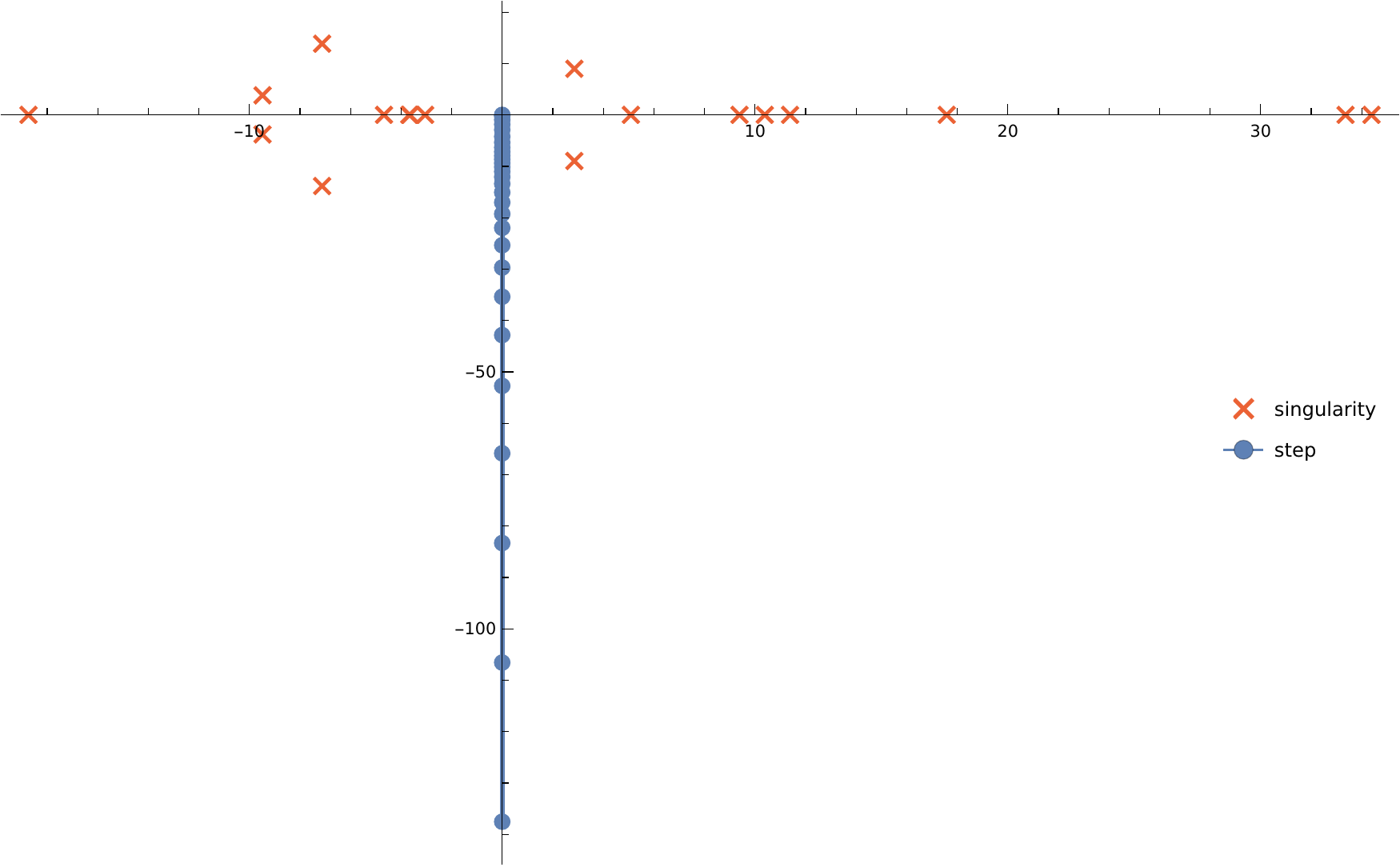}
    \caption{A typical path to solve the differential equation at a certain phase space point.
        This plot shows singularities and steps of the family planar no.\ 1 at $s = (500 \text{~GeV})^{2},\ t = - (300 \text{~GeV})^2$ in the complex plane of $x$.
        The red crosses are the singularities of the differential equation,
        while the blue dots are the steps used to solve the differential equation.
        We approach the origin from $- i \infty$ along the positive imaginary axis.
        The boundary at $x = - i \infty$ is not shown on this plot.}
    \label{fig:equationpath}
\end{figure}

Having fixed all boundary conditions, we are ready to solve the differential equation.
We write the differential equation in terms of a dimensionless variable
\begin{equation}
    x
    =
    \frac{m_{t}^{2} - (173 \text{~GeV})^{2}}{m_{W}^{2}}
    \text{,}
\end{equation}
and make all master integrals dimensionless using $m_{W}^{2}$.
At each phase space point,
we solve the differential equation numerically by moving along the positive imaginary axis from the boundary at $x = - i \infty$ to the physical point $x = 0$.
This involves three steps.
\begin{enumerate}
    \item First, we transform the differential equation at $x = - i \infty$ to a Fuchsian form
        \begin{equation}
            \frac{\partial \boldsymbol{I}}{\partial y}
            =
            \left( \frac{\boldsymbol{A}_{-1}}{y} + \boldsymbol{A}_{0} + \boldsymbol{A}_{1} y + \ldots \right) \boldsymbol{I}
            \text{,}
        \end{equation}
        where $y = 1 / x$ and $\boldsymbol{I}$ is the vector of master integrals.
    \item Second, we use the differential equation to obtain power-logarithmic expansions of the master integrals $\boldsymbol{I}$ in terms of $y$
        in the neighbourhood of $x = - i \infty$ or $y = 0$.
        For each master integral, we write
        \begin{equation}
            \label{equ:series_singular}
            I_{i}
            =
            \sum_{j}^{M} \epsilon^{j}
            \left[
                \sum_{k = 0}^{N} \sum_{l} c_{i j k l} y^{k} \ln^{l} y + \mathcal{O}(y^{N + 1})
            \right]
            +
            \mathcal{O}(\epsilon^{M + 1})
            \text{,}
        \end{equation}
        where the numerical coefficients $c_{i j k l}$ are completely determined by the differential equation and the boundary conditions.
        The parameter $N$ in the above equation is the desired order of the expansion; it controls the truncation error and, eventually, the precision with which $I_i$ is computed. Parameter $M$ is the maximal power of $\epsilon$ in the series.
    \item Finally, using the expansion \eqref{equ:series_singular},
        we move to a regular point $x_{0}$ within the radius of convergence by direct evaluation of $\boldsymbol{I}(y = 1 / x_{0})$.
        Then at each regular point $x_{i}$ along the path of integration,
        we Taylor expand the master integrals by expanding the equation up to order $N$ around $x_{i}$
        \begin{align}
            \frac{\partial \boldsymbol{I}}{\partial x'}
            & =
            \left( \boldsymbol{A}_{0}' + \boldsymbol{A}_{1}' x' + \ldots \right) \boldsymbol{I}
            \text{,}
            \\
            \label{equ:series_regular}
            I_{i}
            & =
            \sum_{j}^{M} \epsilon^{j}
            \left[
                \sum_{k = 0}^{N} c_{i j k} x'^{k} + \mathcal{O}(x'^{N + 1})
            \right]
            +
            \mathcal{O}(\epsilon^{M + 1})
            \text{,}
        \end{align}
        where $x' = x - x_{i}$.
        Once this is accomplished, we move on to the next point $x_{i + 1}$ within the radius of convergence of the new series~\eqref{equ:series_regular}.
        By repeating this expand-evaluate operation, we finally arrive at the physical point $x = 0$.
        Figure~\ref{fig:equationpath} shows a typical situation in the complex $x$-plane.
\end{enumerate}

There are several advantages of this method.
First, it allows us to evaluate master integrals to \emph{arbitrary} precision in reasonable and predictable time,
which can be difficult to do using other numerical methods.
The possibility to increase precision is crucial for a stable evaluation of the amplitude in quasi-singular regions, e.g.\ around thresholds of internal particles.
Second, given an equation and a valid path of analytic continuation in the complex plane, this method produces identical results every time.
This determinism makes the calculation reproducible and allows one to keep numerical errors under control.
Finally, this method is fast enough for practical applications.
The run time depends on the form of the equation, requested precision, depth of the $\epsilon$-expansion, and working precision used in the calculation.
However, for a result accurate to 15 digits, typical run time is about 1--10 seconds per integral, depending on the form of the equation.
Evaluating \emph{all} master integrals at a typical phase space point takes less than an hour on a single CPU core.

We crosschecked the evaluation of master integrals against \texttt{pySecDec}~\cite{Borowka:2017idc,Borowka:2018goh} and \texttt{FIESTA}~\cite{Smirnov:2015mct}.
We also checked the self-consistency of the differential equations by comparing results obtained by integrating along different paths.
Specifically, all master integrals have been checked against \texttt{pySecDec} at an unphysical phase space point
\begin{equation}
    s = - (120 \text{~GeV})^{2}
    \text{,}
    \qquad
    t = - (10 \text{~GeV})^{2}
    \text{,}
\end{equation}
up to the default precision of \texttt{pySecDec} (3--10 digits).
Some of the master integrals have also been checked against \texttt{FIESTA} at another unphysical phase space point
\begin{equation}
    s = 29 \times (80 \text{~GeV})^{2}
    \text{,}
    \qquad
    t = 31 \times (80 \text{~GeV})^{2}
    \text{,}
\end{equation}
up to default \texttt{FIESTA} precision (3--10 digits).

In addition, evaluations at two different phase space points should be connected by a system of differential equations with respect to $s$ and $t$. We pick the following two phase space points,
\begin{align}
    s_{1} & = (160.008 \text{~GeV})^{2}
    \text{,}
    \qquad
    t_{1} = - (80.008 \text{~GeV})^{2}
    \text{,}
    \\
    s_{2} & = (160.032 \text{~GeV})^{2}
    \text{,}
    \qquad
    t_{2} = - (80.032 \text{~GeV})^{2}
    \text{.}
\end{align}
Taking the evaluations at $(s_{2}, t_{2})$ as boundary condition and running the equations from $(s_{2}, t_{2})$ to $(s_{1}, t_{1})$,
we then check against a direct evaluation at $(s_{1}, t_{1})$.
We find that master integrals evaluated at $(s_{1}, t_{1})$ in the two different ways agree
up to the precision used when solving the equations (15 digits in this particular case).

\section{Numerical evaluation}
\label{sec:numeval}

\begin{figure}[ht]
	\centering
	\includegraphics[width=.6\linewidth]{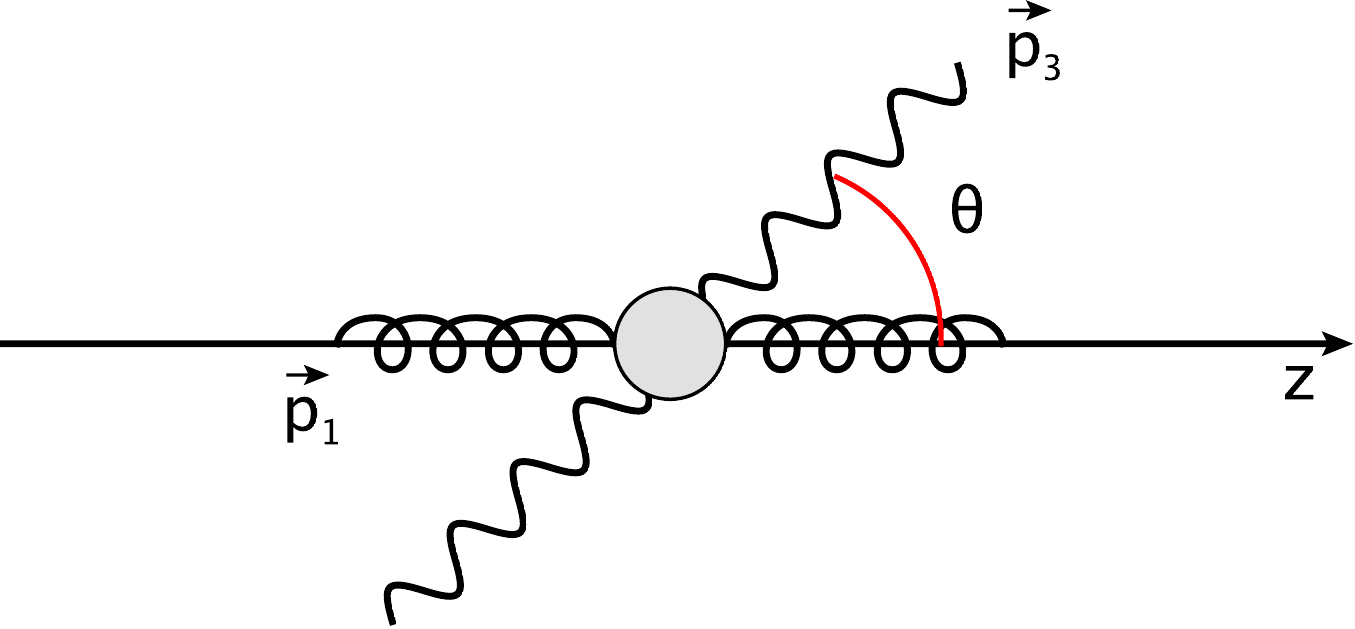}
	\caption{Illustration of the scattering angle $\theta$ between incoming momentum $\vec{p}_1$, parallel to the $z$-axis, and outgoing momentum $\vec{p}_3$.}
	\label{fig:scatteringangle}
\end{figure}

We parametrise the phase space of the $W$ bosons using the angle $\theta$ between $\vec{p}_{1}$ and $\vec{p}_{3}$ in the centre of mass frame, see figure~\ref{fig:scatteringangle},
and the relative velocity of the $W$ boson pair, $\beta$.
These quantities are related to Mandelstam invariants through the following equations
\begin{equation}
    \label{eq:betacostheta}
    s = \frac{4 m_{W}^{2}}{1 - \beta^{2}},
    \qquad
    t = m_{W}^{2} - \frac{s}{2} \left(1 - \beta \cos \theta \right).
\end{equation}

For further references, we present helicity amplitudes evaluated at two phase-space points
\begin{align}
&\textbf{P1:} &\beta &= \frac{1}{2}, & \cos \theta &= - \frac{1}{5}, \label{eq:PS1} \\
&\textbf{P2:} &\beta &= \frac{9}{10}, & \cos \theta &= \frac{4}{5}. \label{eq:PS2}
\end{align}
This corresponds to $\sqrt{s} \approx 185 \text{~GeV}$ and $\theta \approx 102^\circ$ and $\sqrt{s} \approx 367 \text{~GeV}$ and $\theta \approx 37^\circ$ for \textbf{P1} and \textbf{P2} respectively.

To evaluate the tensor structures in~\eqref{eq:ampdecomp} we construct polarisation vectors for the two gluons using spinor-helicity formalism. The polarisation vectors are given by
\begin{align}
\epsilon_{1,L}^\mu = - \frac{1}{\sqrt{2}} \frac{\spba{2}{1}}{[21]},\quad \epsilon_{1,R}^\mu = \frac{1}{\sqrt{2}} \frac{\spab{2}{1}}{\la 21 \ra}, \\
\epsilon_{2,L}^\mu = - \frac{1}{\sqrt{2}} \frac{\spba{1}{2}}{[12]},\quad \epsilon_{2,R}^\mu = \frac{1}{\sqrt{2}} \frac{\spab{1}{2}}{\la 12 \ra}.
\end{align}
Keeping in mind that, eventually, we will be interested in the decay of the $W$ bosons into leptons, we express the polarisation vectors of on-shell $W$ boson states through a current that describes $W^- \to e \overline{\nu}$ and $W^+ \to \overline{e} \nu$ transitions
\begin{align}
\epsilon_{3,L}^\mu = \spab{5}{6},\quad \epsilon_{4,L}^\mu = \spab{7}{8}.
\label{eq:Wcurrents}
\end{align}
The massless momenta are constructed by flattening the massive momenta
\begin{align}
p_5 = p_3 - \frac{m_W^2}{2 p_3 \cdot \eta_1} \eta_1, \quad p_6 = \frac{m_W^2}{2 p_3 \cdot \eta_1} \eta_1, \label{eq:decayp3} \\
p_7 = p_4 - \frac{m_W^2}{2 p_4 \cdot \eta_2} \eta_2, \quad p_8 = \frac{m_W^2}{2 p_4 \cdot \eta_2} \eta_2,\label{eq:decayp4}
\end{align}
where we choose massless reference vectors $\eta_1 = (\sqrt{2},1,1,0)$ and $\eta_2 = (\sqrt{2},1,0,1)$.
The full set of momenta are given in table~\ref{tab:momenta}.

\begin{table}
    \centering
    \begin{tabular}{|c|c|}
        \hline \hline
        \textbf{Phase space point} & \textbf{Momenta} \\
        \hline \hline
        \textbf{P1} &
        $\small{
            \begin{matrix}
                p_{1} & = & ( & 92.37604307, & 0, & 0, & 92.37604307  & )
                \\
                p_{2} & = & ( & 92.37604307, & 0, & 0, & -92.37604307 & )
                \\
                p_{5} & = & ( & 39.37488835, & 7.777358084, & -37.47747591, & -9.237604307 & )
                \\
                p_{6} & = & ( & 53.00115472, & 37.47747591, & 37.47747591, & 0 & )
                \\
                p_{7} & = & ( & 65.22151600, & -64.45598423, & 0, & -9.963545922 & )
                \\
                p_{8} & = & ( & 27.15452707, & 19.20115023, & 0, & 19.20115023 & )
            \end{matrix}}
        $
        \\
        \hline
        \textbf{P2} &
        $\small{
            \begin{matrix}
                p_{1} & = & ( & 183.5325871, & 0,        & 0,        & 183.5325871  & )
                \\
                p_{2} & = & ( & 183.5325871, & 0,        & 0,        & -183.5325871 & )
                \\
                p_{5} & = & ( & 155.3270581, & 79.16327619,  & -19.94432084,  & 132.1434627  & )
                \\
                p_{6} & = & ( & 28.20552903, & 19.94432084, & 19.94432084, & 0 & )
                \\
                p_{7} & = & ( & 174.3120610, & -105.6274936, & 0, & -138.6633593 & )
                \\
                p_{8} & = & ( & 9.220526124, & 6.519896548, & 0, & 6.519896548 & )
            \end{matrix}}
        $
        \\
        \hline \hline
    \end{tabular}
    \caption{Massless incoming and outgoing momenta in units of GeV in the centre of mass frame for the phase space points defined in eq.~\eqref{eq:PS1} and eq.~\eqref{eq:PS2}.}
    \label{tab:momenta}
\end{table}

We label the helicity amplitudes by the helicities of the two incoming gluons and two of the out-going leptons ${\lambda_1 \lambda_2 \lambda_5 \lambda_7}$, where $\lambda_i = L, R$.
$W$ bosons are left-handed and there are four helicity configurations for the gluons.
Two operations allow us to establish relations between these configurations.
First, we can flip all helicities simultaneously by complex conjugation of the polarisation vectors. Second, we can flip the helicities of the $W$ bosons only by swapping the momenta ($p_5 \leftrightarrow p_6$ and $p_7 \leftrightarrow p_8$) in the currents of eq.~\eqref{eq:Wcurrents}.
Hence, only two helicity configurations are independent, we choose to present  ${LLLL}$ and ${LRLL}$.

One-loop helicity amplitudes for the two phase space points defined in eq.~\eqref{eq:PS1} and eq.~\eqref{eq:PS2} are given in table~\ref{tab:evaluation1}.

\begin{table}
	\begin{center}
		\begin{tabular}{|c|c|c|}
			\hline \hline
			$A^{(1)}\vert_{\epsilon=0}$ & \textbf{LLLL} & \textbf{LRLL} \\
			\hline \hline
			\textbf{P1} & \small{$1071.827685027612 + 395.318437150354 i$} & \small{$1711.87290725190 - 4954.09482662664 i$} \\
			\hline
			\textbf{P2} & \small{$7791.28734007197 + 9509.73549766894 i$} & \small{$2134.32524328450 - 2908.70435024589 i$} \\
			\hline \hline
		\end{tabular}
	\end{center}
	\caption{Evaluation of the two independent helicity amplitudes at one loop for the phase space points defined in eq.~\eqref{eq:PS1} and eq.~\eqref{eq:PS2}.}
	\label{tab:evaluation1}
\end{table}

We present $\epsilon$-expansions of the two-loop amplitudes for leading and sub-leading colour in tables~\ref{tab:evaluation2Nc1} and~\ref{tab:evaluation2Nc-1} respectively.
Comparison with the predicted structure of IR poles is also shown.
The renormalisation scale $\mu$ is set to $2 m_W$.
We note that the renormalised two-loop amplitudes~\eqref{eq:A2ren} receive a trivial contribution from the counter term amplitude. It is independent of $N_C$ and we do not include it here.

\begin{landscape}
		\centering
\begin{table}
	\begin{center}
		\begin{tabular}{|c|c|c|c|c|}
			\hline \hline
			\multicolumn{2}{|c|}{\textbf{Two loops}} & \multicolumn{3}{c|}{\textbf{LLLL} } \\
			\hline
			\multicolumn{2}{|c|}{$N_C$} & $\epsilon^{-2}$ & $\epsilon^{-1}$ & $\epsilon^{0}$ \\
			\hline \hline
			\multirow{2}{*}{\textbf{P1}} & $A^{(2),[1]}/A^{(1)}$ & \small{$-0.999999998857788 + 9.6903 \cdot 10^{-11} i$} & \small{$-1.86131749404292 - 4.44620066408116 i$} & \small{$12.61200733077990 - 5.60441510422259 i$} \\
			& IR pole & \small{$-1.000000000000000$} & \small{$-1.86131750171342 - 4.44620066769177 i$} & - \\
			\hline \hline
			\multirow{2}{*}{\textbf{P2}} & $A^{(2),[1]}/A^{(1)}$ & \small{$-1.000000000278483 - 3.35826 \cdot 10^{-10} i$} & \small{$-0.92496050816583 - 4.30331991724938 i$} & \small{$14.3620835041344 + 7.9736182100082 i$} \\
			& IR pole & \small{$-1.000000000000000$} & \small{$-0.92496050665624 - 4.30331991476767 i $} & - \\
			\hline \hline
			\multicolumn{2}{|c|}{\textbf{}} & \multicolumn{3}{c|}{\textbf{LRLL} } \\
			\hline
			\multicolumn{2}{|c|}{$N_C$} & $\epsilon^{-2}$ & $\epsilon^{-1}$ & $\epsilon^{0}$ \\
			\hline \hline
			\multirow{2}{*}{\textbf{P1}} & $A^{(2),[1]}/A^{(1)}$ & \small{$-1.000000002280574 - 1.477331 \cdot 10^{-9} i$} & \small{$-1.50299977076179 - 5.37992305396807 i$} & \small{$13.7636860170288 - 7.2085584481283 i$} \\
			& IR pole & \small{$-1.000000000000000$} & \small{$-1.50299976418128 - 5.37992304294408 i$} & - \\
			\hline \hline
			\multirow{2}{*}{\textbf{P2}} & $A^{(2),[1]}/A^{(1)}$ & \small{$-0.999999992424562 + 2.318144 \cdot 10^{-9} i$} & \small{$1.37986725767052 - 8.54746743169715 i$} & \small{$27.3890551624320 + 3.3867392467224 i$} \\
			& IR pole & \small{$-1.000000000000000$} & \small{$1.37986720742840 - 8.54746746126171 i$} & - \\
			\hline \hline
		\end{tabular}
	\end{center}
	\caption{Evaluation of the two-loop helicity amplitudes, ${LLLL}$ and ${LRLL}$, for the phase space points defined in eqs.~\eqref{eq:PS1} and~\eqref{eq:PS2} for the leading colour contribution. We normalise by the one-loop amplitude $A^{(1)} \vert_{\epsilon=0}$ and show the infrared pole structure for comparison. We set the renormalisation scale $\mu = 2 m_W$.}
	\label{tab:evaluation2Nc1}
\end{table}
\end{landscape}

\begin{table}
	\begin{center}
		\begin{tabular}{|c|c|c|}
			\hline \hline
			\multicolumn{2}{|c|}{\textbf{Two loops}} & \textbf{LLLL} \\
			\hline
			\multicolumn{2}{|c|}{$1/N_C$} & $\epsilon^{0}$ \\
			\hline \hline
			\textbf{P1} & $A^{(2),[-1]}/A^{(1)}$ & $-0.604318842260586 + 0.554150870358548i$ \\
			\hline \hline
			\textbf{P2} & $A^{(2),[-1]}/A^{(1)}$ & $-6.09083779674665 - 6.83926633649785 i$ \\
			\hline \hline
			\multicolumn{2}{|c|}{}& \textbf{LRLL}  \\
			\hline
			\multicolumn{2}{|c|}{$1/ N_C$} & $\epsilon^{0}$ \\
			\hline \hline
			\textbf{P1} & $A^{(2),[-1]}/A^{(1)}$ & $-1.004215354701388 + 0.569698273209762 i$  \\
			\hline \hline
			\textbf{P2} & $A^{(2),[-1]}/A^{(1)}$ & $1.48368538287541 + 1.38326340829964 i$ \\
			\hline \hline
		\end{tabular}
	\end{center}
	\caption{Evaluation of the two-loop helicity amplitudes for the phase space points defined in eqs.~\eqref{eq:PS1} and~\eqref{eq:PS2} for the sub-leading colour contribution, which is finite after mass renormalisation. We normalise by the  one-loop amplitude $A^{(1)} \vert_{\epsilon=0}$ and set the renormalisation scale $\mu = 2 m_W$.}
	\label{tab:evaluation2Nc-1}
\end{table}

\begin{figure}
	\begin{subfigure}{.5\textwidth}
		\centering
		\includegraphics[width=.8\linewidth]{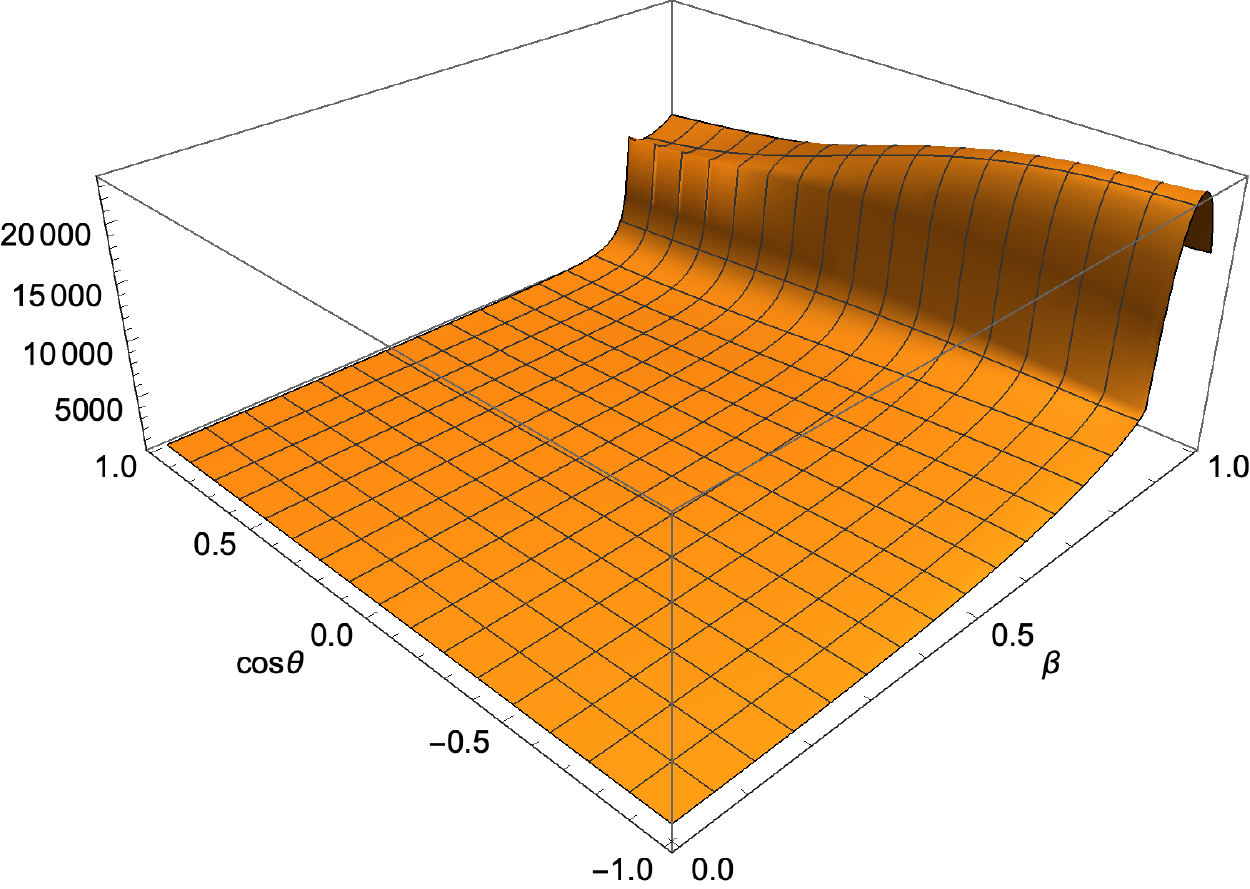}
		\caption{Helicity LLLL}
		\label{fig:sub-first}
	\end{subfigure}
	\begin{subfigure}{.5\textwidth}
		\centering
		\includegraphics[width=.8\linewidth]{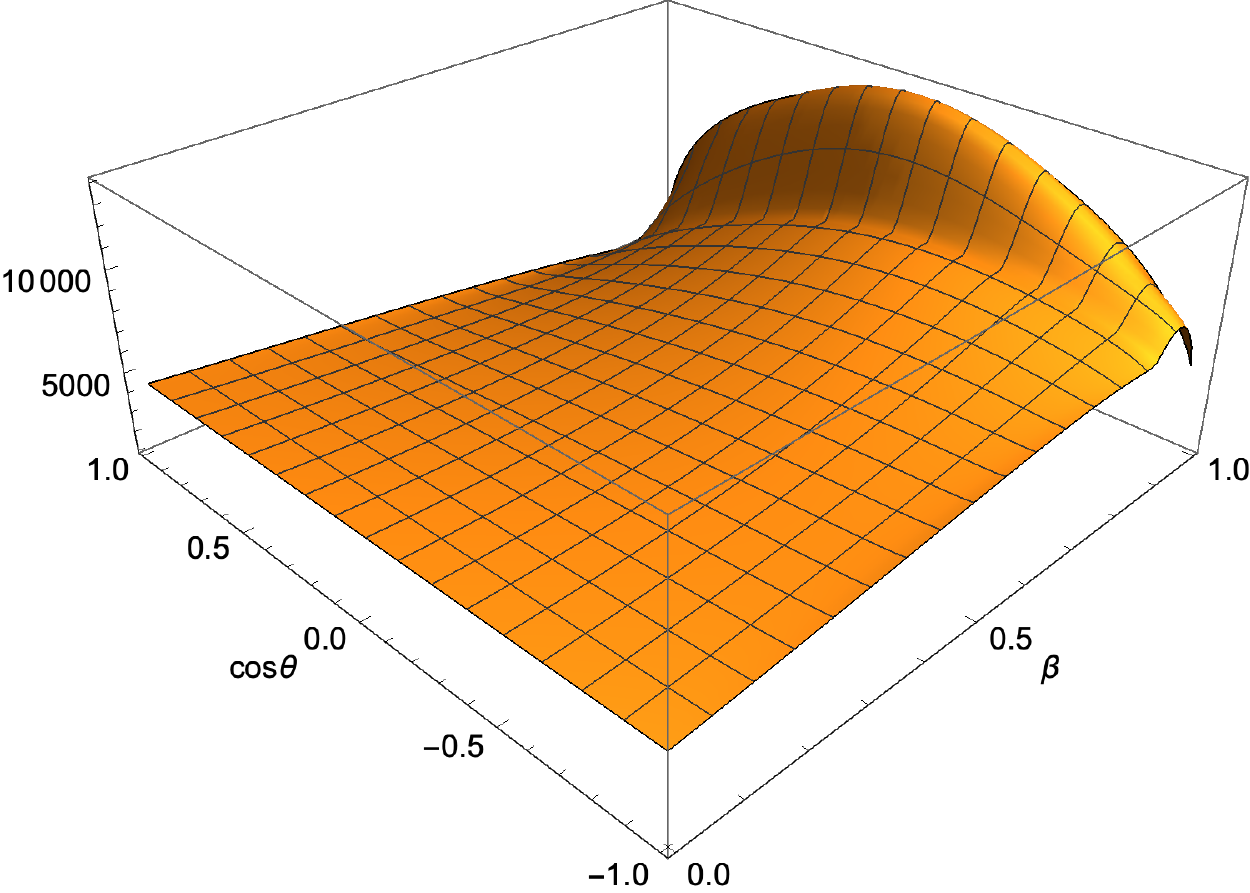}
		\caption{Helicity LRLL}
		\label{fig:sub-second}
	\end{subfigure}
	\caption{Absolute value of the vector-vector plus axial-axial part of the one-loop helicity amplitudes.}
	\label{fig:oneloopgrid}
\end{figure}

For the one-loop amplitudes we construct a uniform, dense 99 by 99 grid in terms of the variables $\beta$ and $\cos \theta$ defined in~\eqref{eq:betacostheta} with step sizes of $0.01$ and $0.02$ in the ranges $[0.01,0.99]$ and $[-0.98,0.98]$ respectively.
The absolute value of the two independent helicity amplitudes are plotted in figure~\ref{fig:oneloopgrid}.
We stress that the helicity amplitudes presented here depend on the polarisation vectors of the on-shell W bosons, see eqs.~\eqref{eq:decayp3} and~\eqref{eq:decayp4}.
To avoid this one can project onto helicity dependent form factors defined in refs.~\cite{Binoth:2006mf,Caola:2015ila,vonManteuffel:2015msa}.

For the two loop amplitude we use a sparse grid for the bulk of phase space, $0.1 \leq \beta < 0.8$.
The step size in $\beta$ is $0.1$ and $\cos \theta$ ranges from $-0.8$ to $0.8$ in steps of $0.2$ with an additional border at $\cos \theta = \pm 0.96$.
For the production threshold, $0.01 \leq \beta < 0.1$ we use a step size of $0.01$ and same resolution for $\cos \theta$ as in the bulk region.
In the high-energy region $0.8 \leq \beta \leq 0.99$ we also use the step size of $0.01$ for $\beta$, but increase resolution in $\cos \theta$ with a step size of $0.04$ in the range from $-0.96$ to $0.96$.

\begin{figure}
	\begin{subfigure}{.5\textwidth}
		\centering
		\includegraphics[width=.8\linewidth]{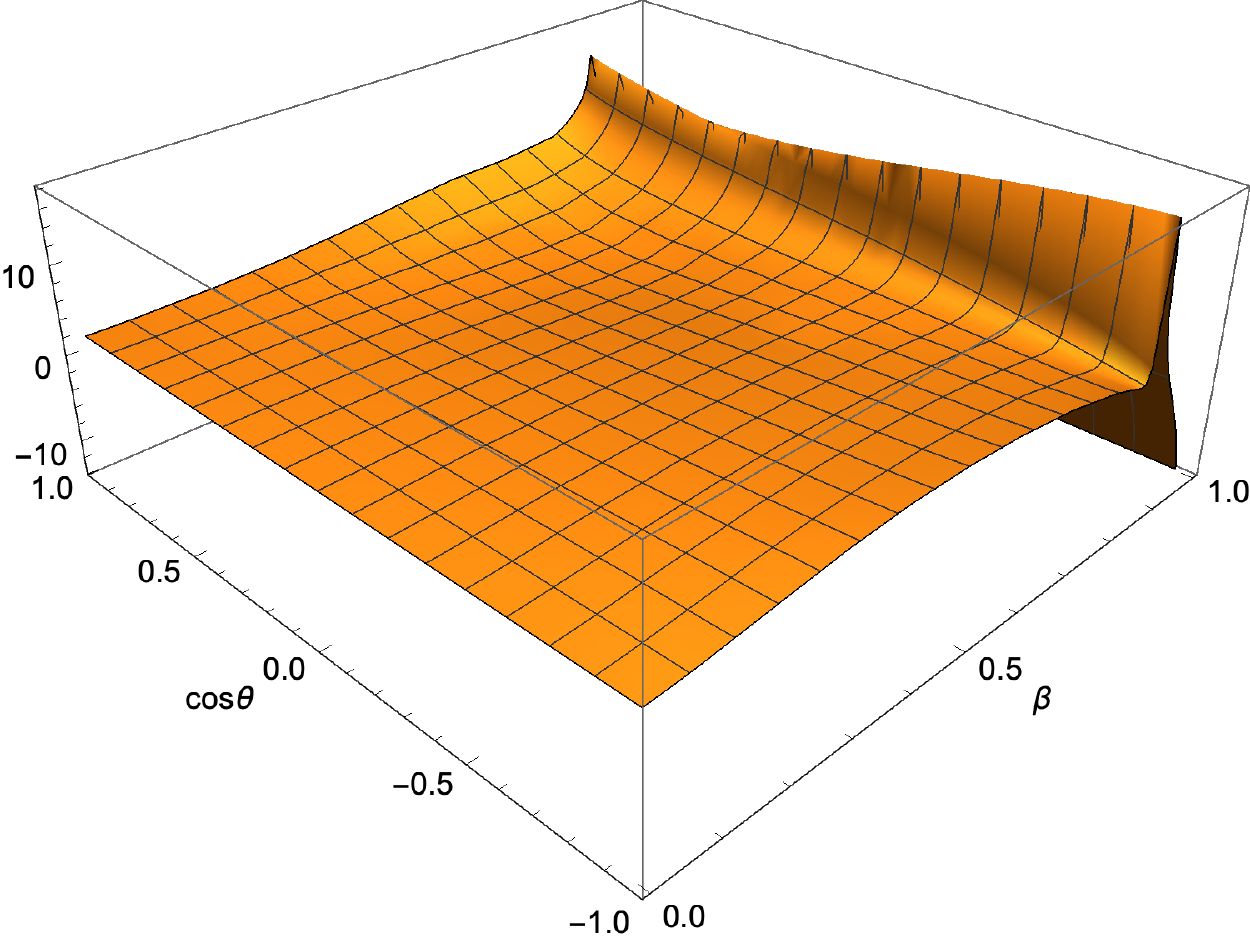}
		\caption{Helicity LLLL}
	\end{subfigure}
	\begin{subfigure}{.5\textwidth}
		\centering
		\includegraphics[width=.8\linewidth]{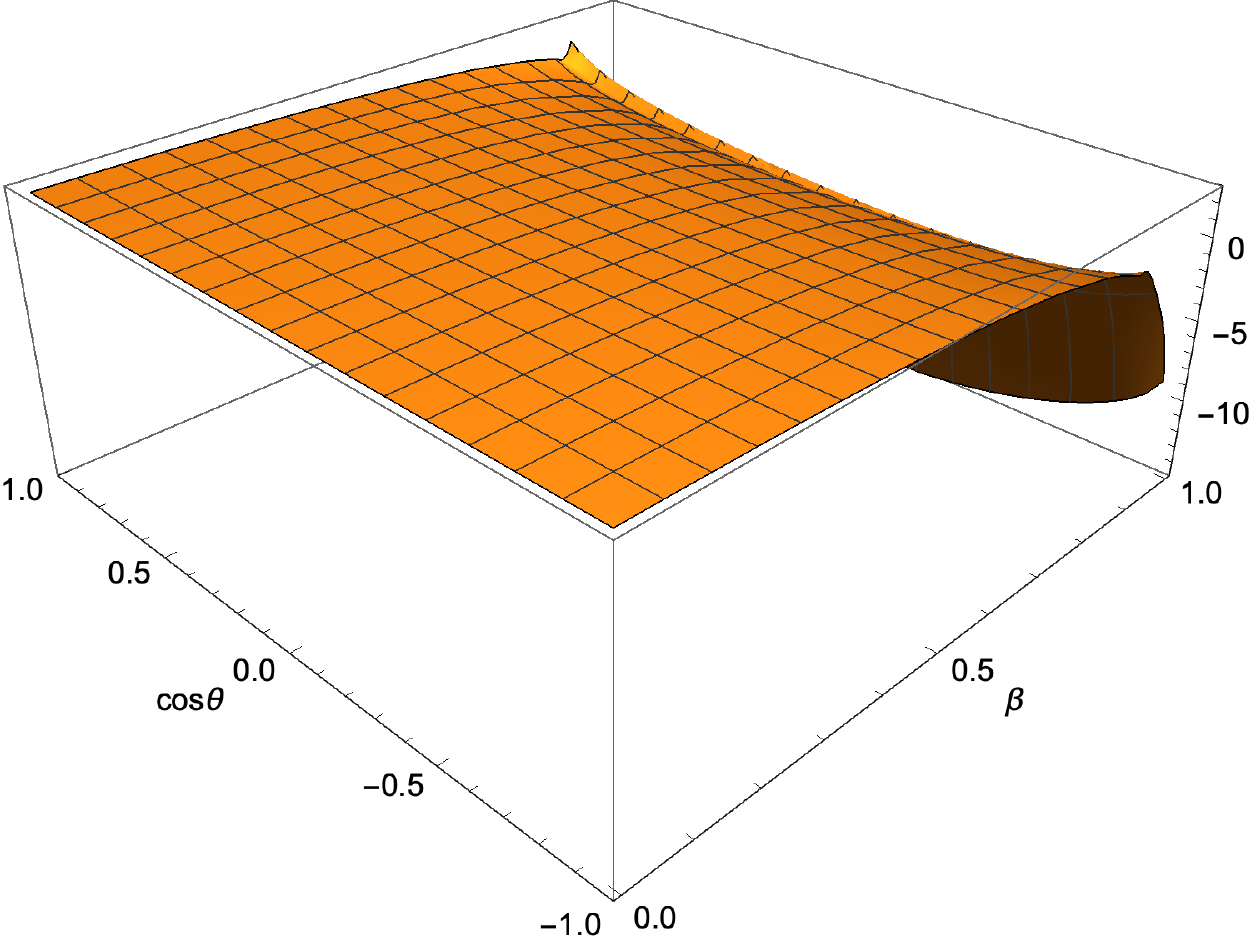}
		\caption{Helicity LRLL}
	\end{subfigure}
	\caption{Finite remainder of the vector-vector and axial-axial part of the two-loop, leading colour ($N_C$), helicity amplitudes interfered and normalised by the leading order amplitude, see eq.~\eqref{eq:interference}. We set the renormalisation scale $\mu = 2 m_W$.}
	\label{fig:twoloopgridNc1}
\end{figure}

\begin{figure}
	\begin{subfigure}{.5\textwidth}
		\centering
		\includegraphics[width=.8\linewidth]{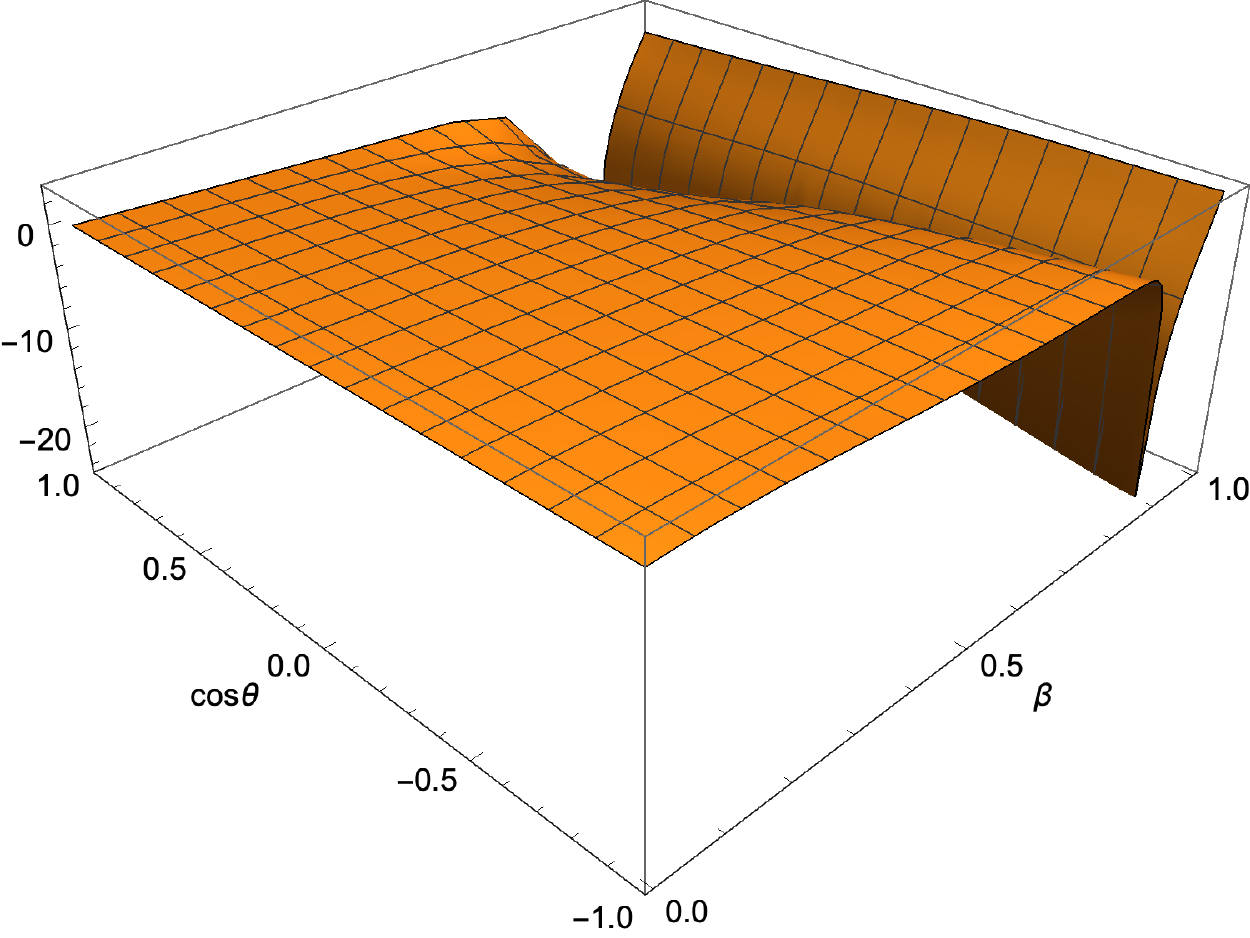}
		\caption{Helicity LLLL}
	\end{subfigure}
	\begin{subfigure}{.5\textwidth}
		\centering
		\includegraphics[width=.8\linewidth]{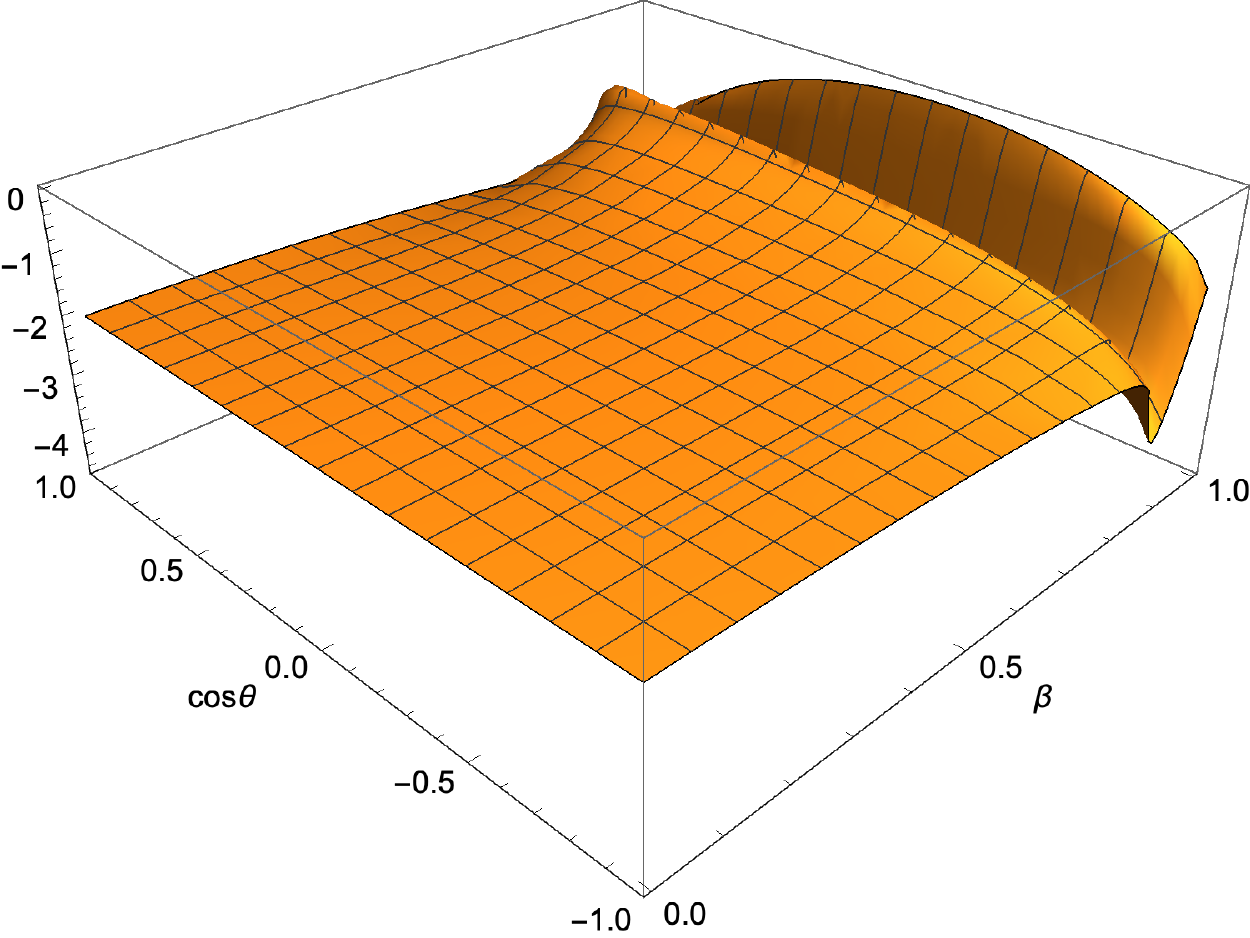}
		\caption{Helicity LRLL}
	\end{subfigure}
	\caption{Finite remainder of the vector-vector and axial-axial part of two-loop, sub-leading in colour ($1 /N_C$), helicity amplitudes interfered and normalised by the leading order amplitude, see eq.~\eqref{eq:interference}. We set the renormalisation scale $\mu = 2 m_W$.}
	\label{fig:twoloopgridNc-1}
\end{figure}

In total 1156 points have been computed to produce plots for the two-loop helicity amplitudes. In figure~\ref{fig:twoloopgridNc1} and~\ref{fig:twoloopgridNc-1} we plot the interference of the finite remainder with the one-loop amplitude,
\begin{align}
\frac{2 \textrm{Re} \left[F^{(2)} A^{(1) \star}  \right]}{\vert A^{(1)} \vert^2}.\label{eq:interference}
\end{align}
For illustration purposes we only show the vector-vector and axial-axial part of the amplitudes in these plots.

\section{Conclusions}\label{sec:conclusions}

In this paper we computed the contribution of the third generation quarks to the two-loop helicity amplitudes for $W$ boson pair production in gluon fusion.
We use projection operators to obtain form factors that can be calculated using integration-by-parts integral reduction.
To overcome the computational bottleneck of the reduction step, we fix the masses of the top quark and the $W$ bosons to integer numbers close to their experimentally determined values.
The master integrals are evaluated numerically by solving a system of differential equations with respect to the top mass parameter.
This approach allows for arbitrary precision and is especially efficient for processes involving massive internal particles.

The present calculation opens up the possibility to include the contribution of third generation quarks into NLO QCD corrections to the cross section of $W$ pair production in gluon fusion.
More generally, this method can be used for numerical calculations of many loop amplitudes with massive particles.
As higher order virtual corrections to many processes involving internal masses are currently beyond the reach of analytic methods, this approach represents an alternative way forward.

\acknowledgments

We are grateful to Kirill Melnikov for guidance and encouragement throughout this project and helpful suggestions for the manuscript.

This research is partially supported by the Deutsche Forschungsgemeinschaft (DFG, German Research Foundation) under grant 396021762 - TRR 257.
Feynman diagrams in figure~\ref{fig:diagrams},~\ref{fig:family}, and~\ref{fig:boundary_example} were generated using \texttt{FeynArts} \cite{Hahn:2000kx}. \texttt{JaxoDraw}~\cite{Binosi:2008ig} was used for the diagrams in table~\ref{tab:colour} and figure~\ref{fig:basictopologies}.

\newpage

\appendix

\section{Integral families}
\label{sec:families}

We define 26 integral families for integral reductions.
There is a single one-loop squared family, 18 planar families labelled planar 1 to 9, each with $p_{1}$ and $p_{2}$ crossed as well.
In the nonplanar case, we define 7 families labelled nonplanar 1 to 5 together with crossed versions of nonplanar 2 and 3.

\begin{table}
    \centering
    \begin{tabular}{|c|c|p{0.7 \linewidth}|}
        \hline \hline
        \multicolumn{2}{|c|}{\textbf{Name}} & \multicolumn{1}{|c|}{\textbf{Definition}} \\
        \hline \hline
        \multicolumn{2}{|c|}{\multirow{2}{*}{one-loop squared}} &
        $
            (l_{2} + p_{3})^{2},
            (l_{2} - p_{1} - p_{2} + p_{3})^{2},
            l_{1}^{2} - m_{t}^{2},
            l_{2}^{2} - m_{t}^{2},
            (l_{1} - p_{1})^{2} - m_{t}^{2},
        $
        \newline
        $
            (l_{1} + p_{2})^{2} - m_{t}^{2},
            l_{1} \cdot l_{2},
            l_{1} \cdot p_{3},
            l_{2} \cdot p_{2}.
        $
        \\
        \hline \hline
        \multirow{18}{*}{planar}
        & \multirow{2}{*}{1} &
        $
            l_{2}^{2},
            (l_{1} + l_{2} - p_{1} + p_{3})^{2},
            l_{1}^{2} - m_{t}^{2},
            (l_{1} - p_{1})^{2} - m_{t}^{2},
            (l_{1} + p_{2})^{2} - m_{t}^{2},
        $
        \newline
        $
            (l_{2} + p_{3})^{2} - m_{t}^{2},
            (l_{2} - p_{1} - p_{2} + p_{3})^{2} - m_{t}^{2},
            l_{1} \cdot p_{3},
            l_{2} \cdot p_{2}.
        $
        \\
        \cline{2-3}
        & \multirow{2}{*}{2} &
        $
            l_{1}^{2},
            (l_{1} - p_{1})^{2},
            (l_{1} + p_{2})^{2},
            (l_{2} + p_{3})^{2},
            (l_{1} + l_{2} - p_{1} + p_{3})^{2},
        $
        \newline
        $
            (l_{2} - p_{1} - p_{2} + p_{3})^{2},
            l_{2}^{2} - m_{t}^{2},
            l_{1} \cdot p_{3},
            l_{2} \cdot p_{2}.
        $
        \\
        \cline{2-3}
        & \multirow{2}{*}{3} &
        $
            l_{1}^{2},
            l_{2}^{2},
            (l_{1} - p_{1})^{2},
            (l_{1} + p_{2})^{2},
            (l_{2} + p_{3})^{2} - m_{t}^{2},
        $
        \newline
        $
            (l_{1} + l_{2} - p_{1} + p_{3})^{2} - m_{t}^{2},
            (l_{2} - p_{1} - p_{2} + p_{3})^{2} - m_{t}^{2},
            l_{1} \cdot p_{3},
            l_{2} \cdot p_{2}.
        $
        \\
        \cline{2-3}
        & \multirow{2}{*}{4} &
        $
            (l_{1} - p_{3})^{2},
            (l_{1} + l_{2} - p_{1} - p_{2})^{2},
            (l_{2} - p_{1} - p_{2} + p_{3})^{2},
            l_{1}^{2} - m_{t}^{2},
            l_{2}^{2} - m_{t}^{2},
        $
        \newline
        $
            (l_{1} - p_{1})^{2} - m_{t}^{2},
            (l_{2} - p_{2})^{2} - m_{t}^{2},
            l_{1} \cdot p_{2},
            l_{2} \cdot p_{3}.
        $
        \\
        \cline{2-3}
        & \multirow{2}{*}{5} &
        $
            l_{1}^{2},
            l_{2}^{2},
            (l_{1} - p_{1})^{2},
            (l_{2} - p_{2})^{2},
            (l_{1} + l_{2} - p_{1} - p_{2})^{2},
        $
        \newline
        $
            (l_{1} - p_{3})^{2} - m_{t}^{2},
            (l_{2} - p_{1} - p_{2} + p_{3})^{2} - m_{t}^{2},
            l_{1} \cdot p_{2},
            l_{2} \cdot p_{3}.
        $
        \\
        \cline{2-3}
        & \multirow{2}{*}{6} &
        $
            l_{2}^{2},
            (l_{2} + p_{2})^{2},
            (l_{1} - p_{3})^{2},
            (l_{1} + l_{2} + p_{2} - p_{3})^{2},
            l_{1}^{2} - m_{t}^{2},
        $
        \newline
        $
            (l_{1} - p_{1})^{2} - m_{t}^{2},
            (l_{2} + p_{1} + p_{2} - p_{3})^{2} - m_{t}^{2},
            l_{1} \cdot p_{2},
            l_{2} \cdot p_{3}.
        $
        \\
        \cline{2-3}
        & \multirow{2}{*}{7} &
        $
            (l_{1} + l_{2} + p_{2})^{2},
            (l_{1} - p_{1} + p_{3})^{2},
            (l_{2} + p_{1} + p_{2} - p_{3})^{2},
            l_{1}^{2} - m_{t}^{2},
            l_{2}^{2} - m_{t}^{2},
        $
        \newline
        $
            (l_{1} - p_{1})^{2} - m_{t}^{2},
            (l_{1} + p_{2})^{2} - m_{t}^{2},
            l_{2} \cdot p_{2},
            l_{2} \cdot p_{3}.
        $
        \\
        \cline{2-3}
        & \multirow{2}{*}{8} &
        $
            l_{1}^{2},
            l_{2}^{2},
            (l_{1} - p_{1})^{2},
            (l_{1} + p_{2})^{2},
            (l_{1} + l_{2} + p_{2})^{2},
        $
        \newline
        $
            (l_{1} - p_{1} + p_{3})^{2} - m_{t}^{2},
            (l_{2} + p_{1} + p_{2} - p_{3})^{2} - m_{t}^{2},
            l_{2} \cdot p_{2},
            l_{2} \cdot p_{3}.
        $
        \\
        \cline{2-3}
        & \multirow{2}{*}{9} &
        $
            (l_{1} + p_{3})^{2},
            (l_{1} + l_{2} + p_{1})^{2},
            (l_{1} - p_{2} + p_{3})^{2},
            (l_{2} + p_{1} + p_{2} - p_{3})^{2},
            l_{1}^{2} - m_{t}^{2},
        $
        \newline
        $
            l_{2}^{2} - m_{t}^{2},
            (l_{1} + p_{1})^{2} - m_{t}^{2},
            l_{2} \cdot p_{2},
            l_{2} \cdot p_{3}.
        $
        \\
        \hline \hline
        \multirow{10}{*}{nonplanar}
        & \multirow{2}{*}{1} &
        $
            l_{1}^{2},
            (l_{1} - p_{1})^{2},
            (l_{1} + p_{2})^{2},
            (l_{2} + p_{3})^{2},
            (l_{1} - l_{2} + p_{2} - p_{3})^{2},
        $
        \newline
        $
            l_{2}^{2} - m_{t}^{2},
            (l_{1} - l_{2} - p_{1})^{2} - m_{t}^{2},
            l_{2} \cdot p_{1},
            l_{2} \cdot p_{2}.
        $
        \\
        \cline{2-3}
        & \multirow{2}{*}{2} &
        $
            l_{2}^{2},
            (l_{2} - p_{2})^{2},
            (l_{1} - p_{3})^{2},
            (l_{1} - l_{2} + p_{2} - p_{3})^{2},
            l_{1}^{2} - m_{t}^{2},
        $
        \newline
        $
            (l_{1} - p_{1})^{2} - m_{t}^{2},
            (l_{1} - l_{2} - p_{1})^{2} - m_{t}^{2},
            l_{2} \cdot p_{1},
            l_{2} \cdot p_{3}.
        $
        \\
        \cline{2-3}
        & \multirow{2}{*}{3} &
        $
            l_{1}^{2},
            l_{2}^{2},
            (l_{1} - p_{1})^{2},
            (l_{2} - p_{2})^{2},
            (l_{1} - l_{2} - p_{1})^{2},
        $
        \newline
        $
            (l_{1} - p_{3})^{2} - m_{t}^{2},
            (l_{1} - l_{2} + p_{2} - p_{3})^{2} - m_{t}^{2},
            l_{2} \cdot p_{1},
            l_{2} \cdot p_{3}.
        $
        \\
        \cline{2-3}
        & \multirow{2}{*}{4} &
        $
            l_{2}^{2},
            (l_{2} - p_{1})^{2},
            (l_{1} + p_{3})^{2},
            (l_{1} - l_{2} + p_{3})^{2},
            (l_{1} - l_{2} - p_{2} + p_{3})^{2},
        $
        \newline
        $
            (l_{1} - p_{1} - p_{2} + p_{3})^{2},
            l_{1}^{2} - m_{t}^{2},
            l_{2} \cdot p_{2},
            l_{2} \cdot p_{3}.
        $
        \\
        \cline{2-3}
        & \multirow{2}{*}{5} &
        $
            l_{1}^{2},
            (l_{1} - l_{2} + p_{3})^{2},
            (l_{1} - l_{2} - p_{2} + p_{3})^{2},
            l_{2}^{2} - m_{t}^{2},
            (l_{2} - p_{1})^{2} - m_{t}^{2},
        $
        \newline
        $
            (l_{1} + p_{3})^{2} - m_{t}^{2},
            (l_{1} - p_{1} - p_{2} + p_{3})^{2} - m_{t}^{2},
            l_{2} \cdot p_{2},
            l_{2} \cdot p_{3}.
        $
        \\
        \hline \hline
    \end{tabular}
    \caption{Definitions of the integral families.
        $l_{1}$ and $l_{2}$ are loop momenta while $p_{1}$, $p_{2}$, and $p_{3}$ are external momenta defined in eq.~\eqref{equ:process}.
    }
    \label{tab:families}
\end{table}

\newpage

\section{Boundary condition of the differential equation}
\label{sec:boundary}

The explicit expressions for the boundary integrals in figure~\ref{fig:boundary} are listed below \cite{tHooft:1978jhc, Chetyrkin:1980pr, Scharf:1993ds, Gehrmann:1999as, Gehrmann:2005pd},
\begin{align}
    I_{1}
    & =
    - \exp(\epsilon \gamma_{E})
    \Gamma(-1 + \epsilon)
    \text{,}
    \\
    I_{2}
    & =
    - \exp(2 \epsilon \gamma_{E})
    \Gamma(-1 + \epsilon)^{2}
    \frac{\Gamma(2 - \epsilon) \Gamma(-1 + 2 \epsilon)}{\Gamma(\epsilon)}
    \text{,}
    \\
    I_{3}(q^{2})
    & =
    \exp(\epsilon \gamma_{E})
    \Gamma(\epsilon) (-1)^{\epsilon} (q^{2})^{-\epsilon}
    \frac{\Gamma(1 - \epsilon)^{2}}{\Gamma(2 - 2 \epsilon)}
    \text{,}
    \\
    I_{4}(q^{2})
    & =
    - \exp(2 \epsilon \gamma_{E})
    \Gamma(-1 + 2 \epsilon) (-1)^{-1 + 2 \epsilon} (q^{2})^{1 - 2 \epsilon}
    \frac{\Gamma(1 - \epsilon)^{3}}{\Gamma(3 - 3 \epsilon)}
    \text{,}
    \\
    I_{5}(q^{2})
    & =
    \exp(2 \epsilon \gamma_{E})
    \Gamma(2 \epsilon) (-1)^{2 \epsilon} (q^{2})^{-2 \epsilon}
    \frac{\Gamma(1 - 2 \epsilon)^{2} \Gamma(1 - \epsilon)^{2} \Gamma(\epsilon)}{\Gamma(2 - 3 \epsilon) \Gamma(2 - 2 \epsilon)}
    \text{,}
    \\
    I_{6}(q^{2})
    & =
    \exp(2 \epsilon \gamma_{E})
    (-1)^{2 + 2 \epsilon} (q^{2})^{-2 - 2 \epsilon}
    \bigg[
        - \frac{\Gamma(1 - \epsilon) \Gamma(1 + \epsilon) \Gamma(1 - 2 \epsilon)^{4} \Gamma(1 + 2 \epsilon)^{3}}{\epsilon^{4} \Gamma(1 - 4 \epsilon)^{2} \Gamma(1 + 4 \epsilon)}
        \nonumber \\
        & + \frac{\Gamma(1 - \epsilon)^{2} \Gamma(1 + \epsilon) \Gamma(1 - 2 \epsilon) \Gamma(1 + 2 \epsilon)}{2 \epsilon^{4} \Gamma(1 - 3 \epsilon)}
        {}_{3}F_{2}(1, -4 \epsilon, -2 \epsilon; 1 - 3 \epsilon, 1 - 2 \epsilon; 1)
        \nonumber \\
        & + \frac{- 4 \Gamma(1 - \epsilon)^{2} \Gamma(1 - 2 \epsilon) \Gamma(1 + 2 \epsilon)}{\epsilon^{2} (1 + \epsilon) (1 + 2 \epsilon) \Gamma(1 - 4 \epsilon)}
        {}_{3}F_{2}(1, 1, 1 + 2 \epsilon; 2 + \epsilon, 2 + 2 \epsilon; 1)
        \nonumber \\
        & + \frac{- \Gamma(1 - \epsilon)^{3} \Gamma(1 + 2 \epsilon)}{2 \epsilon^{4} \Gamma(1 - 3 \epsilon)}
        {}_{4}F_{3}(1, 1 - \epsilon, -4 \epsilon, -2 \epsilon; 1 - 3 \epsilon, 1 - 2 \epsilon, 1 - 2 \epsilon; 1)
    \bigg]
    \text{,}
\end{align}
where $q^{2}$ corresponds to the four-momentum squared of the external legs.
Note that $I_{1}$ and $I_{3}$ are one-loop integrals,
thus they enter the boundary condition through the products among themselves.

\newpage

\bibliographystyle{JHEP}
\bibliography{references}

\end{document}